\documentclass{raa_twocolumn} 

\usepackage{amsmath}	
\usepackage{amssymb}	
\usepackage[pagebackref=true]{hyperref}
\usepackage{pifont}
\usepackage{marvosym}
\usepackage{graphicx}
\usepackage{times}
\usepackage{txfonts}
\usepackage{natbib}
\usepackage{multirow}
\usepackage{enumitem}
\usepackage{url}
\usepackage{color}
\usepackage{xcolor}
\usepackage{orcidlink}
\bibpunct{(}{)}{;}{a}{}{,}
\hypersetup{
    colorlinks=true,
    linkcolor=blue,
    filecolor=blue,      
    urlcolor=blue,
    citecolor=blue
}
\begin{document}


\title{Disentangling Multi-epoch Single-lined Binary Star Spectra to Reveal Interstellar Medium Features via MCMC}

\volnopage{Vol.0 (202x) No.0, 000--000}      
\setcounter{page}{1}          

\author{
Shao-Song Huang \inst{1,2}\orcidlink{0009-0003-7705-6846} \and
Jian-Jun Chen \inst{1,2}\orcidlink{0000-0003-4525-1287} \and
Xiao-Xiao Ma \inst{1,2}\orcidlink{0000-0002-9279-2783} \and
Ji-Wei Shi \inst{1}\orcidlink{0009-0001-4516-8486}
}

\institute{
National Astronomical Observatories,Chinese Academy of Sciences, Beijing 100101, China\\
\and
School of Astronomy and Space Science, University of Chinese Academy of Sciences, Beijing 100049, China,
\\
e-mail: \href{mailto:jjchen@nao.cas.cn}
{jjchen@nao.cas.cn}
\\
\vs\no
   {\small Received 202x month day; accepted 202x month day}}

\abstract{Diffuse Interstellar Bands (DIBs) are interstellar absorption features with unidentified carriers in the optical and near-infrared spectra. When observed in cool star spectra, they are heavily contaminated by stellar lines. Traditional methods rely on template matching to remove the stellar component, but their accuracy is fundamentally limited by template quality. We present a template-independent algorithm that separates stellar and interstellar features in binary star spectra by exploiting orbital motion: stellar lines shift with changing radial velocity (RV), whereas DIBs remain stationary in the rest frame. The MCMC-RVSD algorithm, implemented as an RV-based Bayesian Markov Chain Monte Carlo framework, fits both components simultaneously. We validated our approach by applying it to mock spectra simulating LAMOST medium-resolution observations. The method precisely recovers the parameters of DIB $\lambda\,6614$, achieving a wavelength precision of $0.01\,\mathrm{\AA}$. With the adopted empirical priors, it performs reliably for spectra with $R = 3000$--$7500$. We applied the method to 434 real binary sources from the LAMOST medium-resolution survey and introduced the MCMC autocorrelation time ($\tau_{\mathrm{DIB}}$) to assess the fit quality. Our analysis shows that an RV span $> 20\,\mathrm{km\,s^{-1}}$ is the primary driver of successful line separation. Denser orbital sampling (smaller RV gap) provides further improvement, and the algorithm remains robust against typical observational noise at $\mathrm{SNR} > 50$ and RV errors $< 5\,\mathrm{km\,s^{-1}}$. We compiled a high-precision catalog of DIB $\lambda\,6614$ measurements for these 434 systems. Under ideal conditions, our approach yields significantly smaller parameter uncertainties than traditional template matching. Additionally, the MCMC-RVSD algorithm can also handle multi-stellar line contamination, non-Gaussian DIBs and other DIBs by adjust model and prior settings. This algorithm provides a reliable, general-purpose tool for precise measurement of interstellar features in time-domain spectroscopic surveys.
}

\authorrunning{Huang, S.-S. et al. }            
\titlerunning{Disentangling Multi-epoch SB1 Binary Spectra for ISM Features} 

\maketitle

%
%


\section{Introduction}           
\label{sect:intro}

The interstellar medium (ISM) is the gas and dust in interstellar space. Its absorption lines serve as powerful tools for tracing different gas phases and probing ISM properties. For instance, the Na I D doublet is typically used to trace cold and dense neutral gas, while the Ca II H \& K lines are used to probe the warmer and partially ionized medium. UV absorption lines, such as Fe II and Mg II, are widely used to investigate the interstellar and circumstellar media (\citealt{2002ApJS..139..439R, 1993A&A...278..549B}). 

Among these lines, Diffuse Interstellar Bands (DIBs), unidentified absorption lines at optical and near-infrared wavelengths, present unique characteristics and remain a longstanding mystery in astrophysics. Unlike atomic lines, DIBs usually exhibit weak absorptions, broader profiles, uncertain central wavelengths, and numerous unidentified carriers (\citealt{1995ARA&A..33...19H}). Since Heger first discovered DIBs in the early 1920s, more than 600 DIBs have been found, yet their carriers remain almost entirely unknown (\citealt{1922LicOB..10..146H, 2019ApJ...878..151F, 2022ApJS..262....2H, 2026A&A...713A.177M}). The detection of fine substructures, such as the triple-peak structure of $\lambda$6614, suggests that DIB carriers are likely large molecules (\citealt{1996A&A...307L..25E, 2004ApJ...611L.113C}) . In the history of the search for carriers, $C_{60}^{+}$ broke the long-standing mystery. Combined observational and laboratory efforts have confirmed that several DIBs originate from fullerene ions (\citealt{2019ApJ...875L..28C}).

To identify additional carriers, obtaining precise line parameters (intrinsic profiles and rest-frame central wavelengths) is essential. Since DIBs are usually seen on stellar spectra, the star's own absorptions make them difficult to clearly measure. Over the past few decades, various techniques have been developed to effectively separate the stellar and DIB components, each with its own unique advantages and limitations.

The most widely adopted approach is the standard star technique (or the pair method), in which the target spectrum is divided by (or directly compares it with) that of an unreddened ``standard" star of the same spectral type, so that the stellar lines cancel out (\citealt{1975ApJ...196..129H, 2011ApJ...727...33F, 2019ApJ...878..151F}). Because a close spectral match is required, this technique is most straightforward to apply to hot stars, whose smooth continua and sparse metal lines leave little residual contamination. For cool stars, their crowded metal or molecular lines make finding a single perfect 'twin' impractical, but the same idea has been successfully extended through template matching or template subtraction. The comparison spectrum is an empirical template co-added from many low-reddening stars of similar atmospheric parameters and is then divided out of (or subtracted from) the target spectrum to isolate the interstellar features (\citealt{2012MNRAS.425.1763Y, 2013ApJ...778...86K, 2024A&A...691A.282M, 2025A&A...702A..58S}).

To overcome the scarcity of standard stars, the synthetic template method was introduced. This technique constructs template spectra using stellar atmosphere models (e.g., ATLAS9, DETAIL) (\citealt{2022A&A...662A..81E}), allowing generation of noise-free templates for any stellar parameter combination. However, its accuracy is strictly limited by the completeness of atomic line lists and the precision of the physical models; missing lines in the model lead to false DIB detections.

When observing O-type and B-type stars, measuring certain DIBs becomes significantly simpler. Unlike cooler spectral types (types A, F, G, and K), these hot stars have smoother continua with very few metallic absorption lines, which naturally reduces stellar contamination in the DIB regions (\citealt{2025RAA....25i7001M}). Instead of relying on theoretical models, this method verifies the presence of DIBs by directly comparing the target spectrum side-by-side with an unreddened standard star of the same spectral type. For DIBs that are almost free from stellar contamination, such as DIB $\lambda$6196 and $\lambda$6614, line parameters can be obtained directly through Gaussian fitting (\citealt{2010ApJ...708.1628M}). However, since O-type and B-type stars are rare, this method is severely limited by the small sample size.

Recently, median filtering has emerged as a novel method for measuring DIBs. It operates by stacking spectra to suppress common components, thereby isolating DIB features. \cite{2025MNRAS.543.1429C} employs composite spectroscopy (\citealt{2014MNRAS.439.3139Z, 2018ApJ...866...36L}) to group quasar spectra by identical Mg II absorption strength, measures DIB $\lambda 4430$ via median filtering, and achieves measurement uncertainties at the m\AA \ level. \cite{2026A&A...713A.177M} employs the UNWIND to separate stellar and interstellar components in a Galactic massive binary via median filtering, and constructs the most detailed catalog of 631 DIBs to date (116 new). The application of median filtering to cool star measurements remains unexplored, representing a promising new direction for future investigation.

In summary, while existing methods provide valuable insights into DIB measurements for cool stars, their reliance on static spectra or theoretical templates limits their precision and broad applicability. To overcome these limitations, we propose a novel, template-independent method that leverages orbital dynamics to cleanly separate stellar and interstellar components.

Unlike the methods above, which rely on static spectra, binary systems provide a unique and dynamic perspective. In Single-Lined Spectroscopic Binaries (SB1), the stellar absorption lines shift due to orbital motion (Doppler shift), whereas the DIBs, originating from the intervening ISM, remain stationary in the observer's frame. In this study, we employ the star's radial velocity as the key parameter in a Markov Chain Monte Carlo (MCMC) framework to mathematically separate the moving stellar component from the stationary DIBs. We refer to this approach as the MCMC Radial-Velocity Spectral Decomposition(MCMC-RVSD) method. Unlike template matching, this method does not require template spectra, so it avoids errors from mismatched templates. By using the star's orbital information effectively, we can obtain precise parameters for the DIB lines.

\section{Mock Spectra Construction}
\label{sec:maths} 

To verify the feasibility of the MCMC-RVSD method, it was necessary to construct mock spectra for testing. The construction process consists of three steps: establishing the spectral framework, adding spectral lines, and injecting noise. 

\subsection{Spectral Framework}
We selected the red-end spectra from the Large Sky Area Multi-Object Fiber Spectroscopy Telescope medium-resolution survey (LAMOST MRS) DR13 as our framework. Specifically we simulated the DIB $\lambda$6614 feature and a neighboring stellar line. The parameters of the mock spectra were consistent with the LAMOST MRS red-end spectra: a wavelength coverage of 6300–6800 $\rm{\AA}$, a resolution of 7500, and a sampling step of 0.12 $\rm{\AA}$ (\citealt{2012RAA....12.1197C,2012RAA....12.1243L,2012RAA....12..723Z}). Additionally, we assumed a signal-to-noise ratio (SNR) of 200, with the continuum normalized to unity.

\subsection{Spectral Lines}
Next, we injected spectral lines in this framework. This simulation considered only one stellar absorption line and one DIB profile. We modeled both features using Gaussian profiles, where the combined normalized flux $F$ is defined as

\begin{equation} 
F(\lambda) = F_{\text{star}}(\lambda) \cdot F_{\text{DIB}}(\lambda) \label{eq:FF} 
\end{equation}

Where $A$, $\lambda$ and $\sigma$ denote the absorption depth, central wavelength and Gaussian width, respectively. The profiles are described by

\begin{equation} F_{\text{star}}(\lambda) = 1 - A_{\text{star}} \cdot \exp\left(-\frac{(\lambda - \lambda_{\text{shift}})^2}{2\sigma_{\text{star}}^2}\right) \label{eq:FSTAR} \end{equation}

\begin{equation} F_{\text{DIB}}(\lambda) = 1 - A_{\text{DIB}} \cdot \exp\left(-\frac{(\lambda - \lambda_{\text{DIB}})^2}{2\sigma_{\text{DIB}}^2}\right) \label{eq:FDIB} \end{equation}

The central wavelength of the DIB is fixed in the rest frame of the interstellar medium, while the stellar line is Doppler-shifted according to the radial velocity (RV):

\begin{equation}
\lambda_{\text{shift}} = \lambda_{\text{star}} \left(1 + \frac{\text{RV}}{c}\right)
\label{eq:RV}
\end{equation}

To ensure the simulation reflects realistic observational conditions, the parameters for the two spectral lines were carefully selected based on empirical data. The DIB $\lambda$6614 parameters are derived from the mean values of LAMOST MRS template matching measurements, representing a typical absorption profile (\citealt{2025A&A...702A..58S}). Specifically, the parameters for the DIB are set as $\lambda_{\text{DIB}} = 6613.709\,\rm{\AA}$, $A_{\text{DIB}} = 0.049$, and $\sigma_{\text{DIB}} = 0.546\,\rm{\AA}$. To identify potential stellar contamination, we initially queried the NIST atomic database\footnote{\url{https://www.nist.gov/pml/atomic-spectra-database}} around $6614\,\rm{\AA}$. The search identified the Fe~I line at $6613.825\,\rm{\AA}$ as the dominant candidate, followed by Th~I lines ($6613.373\,\rm{\AA}$, $6613.397\,\rm{\AA}$), a Y~II line ($6613.731\,\rm{\AA}$) and a Co~II line ($6614.780\,\rm{\AA}$) (\citealt{NIST_ASD}). To validate this selection in an astrophysical context, we examined the spectrum of G2V-type stars (solar analogs) using the SpectroWeb 2.0 database\footnote{\url{http://spectra.freeshell.org/spectroweb.html}} (\citealt{1966sst..book.....M,2011CaJPh..89..395L}). This analysis confirmed that the Fe~I line is the most prominent feature in the target window. Consequently, we fitted this line on the website and degraded the resolution from $350,000$ to $7500$, resulting in $\lambda_{\text{star}} = 6613.825\,\rm{\AA}$, $A_{\text{star}} = 0.011$, and $\sigma_{\text{star}} = 0.380\,\rm{\AA}$.

The RV variability was modeled based on the binary star G17351752086699. We utilized the RV values and RV errors from 24 LAMOST epochs as the ``true'' values for constructing the spectra. These values were adopted from the LAMOST $rv_{r0}$, which represents radial velocities derived from red-end spectra via cross-correlation with theoretical templates, while the associated errors ($rv_{r0\_err}$) were estimated based on the SNR and the chi-square of the template match (\citealt{2019ApJS..244...27W}). To generate the ``observed" RV inputs for the MCMC-RVSD algorithm, we perturbed these true values by adding Gaussian noise with standard deviations equal to their measurement errors.

\subsection{Noise}

Observational noise degrades the SNR of spectra, thereby influencing the resolution capability of the algorithm. Common noise sources include photon noise, readout noise, sky background, dark current, among others. In this simulation, to treat the SNR as a single controllable parameter, we adopted a simplified assumption: the combined effect of all noise sources on the spectrum can be modeled by a Gaussian noise model directly linked to the target SNR.

According to the LAMOST medium resolution data description documentation\footnote{\url{https://www.lamost.org/dr12/v1.0/doc/mr-data-production-description}}, the provided SNR represents the average over all pixels. The pipeline supplies the inverse variance ($\chi$) for each pixel, from which the flux error ($f_{\text{err}}$) and SNR are derived as:

\begin{equation}
\chi = \frac{1}{f_{\text{err}}^2}
\label{eq:chi-ferr}
\end{equation}

\begin{equation} 
\text{SNR} = f \cdot \sqrt{\chi} 
\label{eq:snr-chi} 
\end{equation}

where $f$ is the flux in the pixel. To directly control the noise level via the target SNR, we combined the Gaussian noise into the mock spectra based on the SNR:

\begin{equation}
\text{Noise}\sim N(0, {(\frac{f}{\text{SNR}})}^2)
\label{eq:demo2}
\end{equation}

Consequently, the corresponding flux error array for the mock data is:

\begin{equation}
f_{\text{err}}=\frac{f}{\text{SNR}}
\end{equation}

This approach effectively consolidated the combined effects of various noise sources into a single parameter (SNR) for controlled testing of the algorithm's robustness.

\begin{figure}[ht]
    \begin{minipage}[t]{0.995\linewidth}  
    \centering
    \includegraphics[width=1\linewidth]{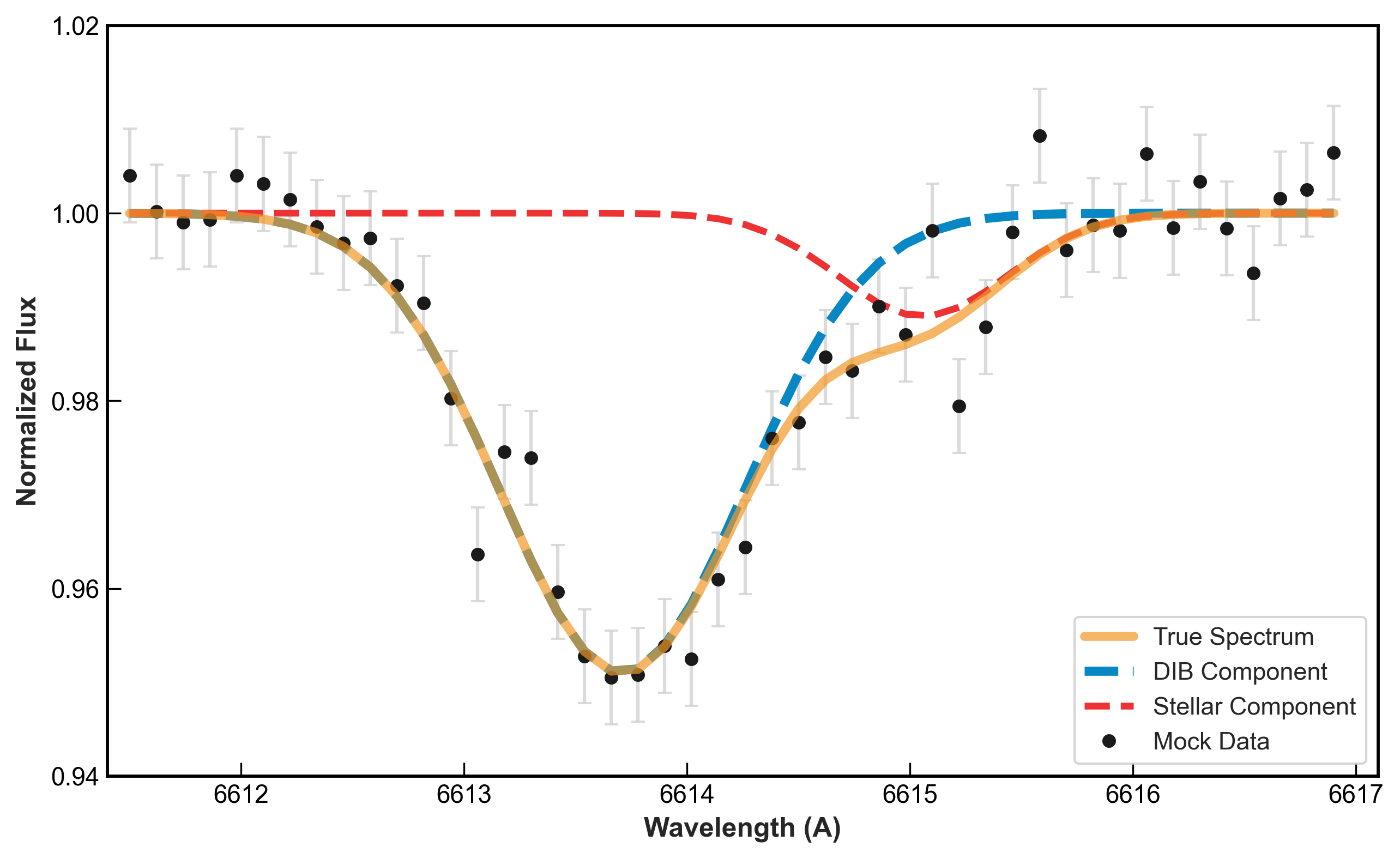} 
    \caption{\label{fig:mock_spectra}\small Decomposition of a mock SB1 spectrum with RV = 55.85 km/s. The mock data (black dots) are fitted with a theoretical model (yellow solid line), consisting of DIB $\lambda$6614 (blue dotted line) and a stellar Fe I line at 6613.825 \rm{\AA} (red dashed line).}
    
    \end{minipage}
\end{figure}

Selecting an appropriate wavelength range could effectively minimize interference from other signals and improve the fitting performance. Centering on the DIB $\lambda$6614 feature, the spectra were trimmed to a range of 6611.5--6617.0~\AA{} and subsequently resampled at an interval of 0.12~\AA. The final mock spectrum is presented in Figure~\ref{fig:mock_spectra}. The figure displays a mock spectrum with an RV of $55.85$~km~s$^{-1}$. The presence of stellar lines is shown to significantly alter the profile of the DIB. Furthermore, the data points with error bars closely align with the theoretical spectrum, while their vertical scatter illustrates the effect of the introduced Gaussian noise.

\section{Bayesian Analysis}
\label{sec:Bayesian} 

\subsection{Bayesian Inference Framework}
Due to orbital motion, the stellar spectral lines in binary stars shift according to the changing RV, while the overlying DIB feature remains stationary, as it is unrelated to the binary orbital motion.

To constrain the parameters of the stellar absorption line and the DIB, we employed a robust Bayesian inference framework, utilizing the varying RV values to solve for the moving stellar line and the stationary DIB line. A physical model was constructed based on Equations~\eqref{eq:FF}--\eqref{eq:RV}, which incorporated six free parameters denoted collectively as $\theta$: $
\lambda_{\text{star}},\ A_{\text{star}},\ \sigma_{\text{star}},\ \lambda_{\text{DIB}},\ A_{\text{DIB}},\ \sigma_{\text{DIB}}
$. These are the central wavelength, absorption depth, and line width for the stellar line and the DIB, respectively.

According to Bayes' theorem, the posterior distribution $P(\theta  |\text{Data})$ is proportional to the likelihood function $P(\text{Data}|\theta)$ times the prior distribution $P(\theta)$: 

\begin{equation}
\label{eq:dataP}
P(\theta  |\text{Data})\propto P(\text{Data}|\theta)P(\theta)
\end{equation}

\subsection{Likelihood Function with Effective Variance and SNR Weighting}
\label{subsec:Likelihood}
Assuming the observational noise follows a Gaussian distribution, the log-likelihood function is expressed as the sum of squared residuals weighted by the variance. However, standard weighting only accounts for flux error ($\sigma_{\text{flux}}$). In our analysis, the RV used to shift the stellar lines to the observed frame also carries error ($\sigma_{\text{RV}}$). Ignoring $\sigma_{\text{RV}}$ will lead to an underestimation of fitting errors.

However, accurately accounting for $\sigma_{\text{RV}}$ is non-trivial, as these errors propagate nonlinearly into the parameter posteriors. To address this issue, we employed the effective variance method (\citealt[Chapter~15]{2007nras.book.....P}), which projected $\sigma_{\text{RV}}$ onto $\sigma_{\text{flux}}$ by incorporating it into the variance. Specifically, the $\sigma_{\text{RV}}$ was first converted into a wavelength error via the Doppler formula. This wavelength error was then propagated to the flux scale using a first-order Taylor expansion of the spectrum. The local gradient of the spectrum acted as the sensitivity coefficient: in regions with steep slopes, such as line wings, even a small wavelength shift caused by $\sigma_{\text{RV}}$ can lead to a large change in flux. By projecting the wavelength error using this gradient, we calculate the propagated flux variance component, denoted as $\sigma_{\text{RV} \to \text{flux}}^2$. The total effective variance $\sigma_{\text{eff}, i}^2$ for the $i$-th data point is given by

\begin{equation}
\sigma_{\text{eff}, i}^2 = \sigma_{\text{flux}, i}^2 + \sigma_{\text{RV} \to \text{flux}, i}^2
\end{equation}

where $\sigma_{\text{flux}, i}$ is the observational flux noise. $\sigma_{\text{RV} \to \text{flux}}^2$ is explicitly defined as

\begin{equation}
\sigma_{\text{RV} \to \text{flux}, i}^2 = \left( \frac{\partial F_{\text{mod}}(\lambda_i)}{\partial \lambda} \right)^2 \left( \frac{\lambda_i}{c} \sigma_{\text{RV}} \right)^2
\end{equation}

Here, $c$ is the speed of light, and $\sigma_{\text{RV}}$ is the RV error. The term $(\lambda_i/c)\sigma_{\text{RV}}$ represents the wavelength uncertainty induced by the $\sigma_{\text{RV}}$ via the Doppler formula, while $\partial F_{\text{mod}}/\partial \lambda$ is the slope of the spectrum at wavelength $\lambda_i$, representing the local spectral gradient.

Furthermore, when jointly fitting multiple spectra of the same target source, the individual observations often have varying data qualities. To optimally combine them, we introduced an SNR weighting scheme (i.e., variance scaling). We define a normalized weight for the $k$-th spectrum as

\begin{equation}
w_k = \frac{\text{SNR}_k}  {\langle \text{SNR} \rangle}
\end{equation}

where $\langle \text{SNR} \rangle$ is the mean SNR across all spectra of the target. The effective variance is inversely scaled by this weight to become

\begin{equation}
\tilde{\sigma}_{\text{eff}, k,i}^2 = \sigma_{\text{eff}, k,i}^2 / w_k
\end{equation}

Consequently, spectra with higher SNR are assigned smaller effective variances, enhancing their contribution. The final log-likelihood function incorporating both RV-propagated variance and SNR weighting is formulated as

\begin{equation}
\begin{split}
\ln \text{L}(\boldsymbol{\theta}) =
& -\frac{1}{2} \sum_{k=1}^{N_{\text{spec}}} \sum_{i=1}^{N_{\text{pix}}} \\
& \left[ \frac{(F_{\text{obs}, k,i} - F_{\text{mod}, k,i} (\boldsymbol{\theta}))^2}{\sigma_{\text{eff}, k,i}^2 / w_k} + \ln\left(2\pi \frac{\sigma_{\text{eff}, k,i}^2}{w_k}\right) \right]
\end{split}
\label{eq:likelihood snr}
\end{equation}

In the propagation of $\sigma_{\text{RV}}$, we approximate the nonlinear mapping between flux and wavelength using a first-order Taylor expansion. However, this linear approximation may introduce systematic error, particularly at extremely high SNR and/or large $\sigma_{\text{RV}}$ (e.g., $\sigma_{\text{RV}}$ = 10~km~s$^{-1}$ or SNR = 20,000). This systematic error will be amplified by these parameters, potentially causing the true value to fall outside the $3\sigma$ range of the posterior distribution. In Section~\ref{subsec:Radial Velocity}, we will further analyze the impact of $\sigma_{\text{RV}}$.


\subsection{Priors and MCMC Sampling}
The design of prior information influences the convergence behavior of MCMC. For the spectral parameters, uniform priors were adopted as the most conservative choice, given insufficient information to justify Gaussian distributions.

\begin{table*}[!htb]
  \caption[]{\small Summary of model parameters: priors, initial values, and posterior constraints}
  \label{tab:full_fit_results}
  \begin{center}
  \begin{tabular}{c|c c c r r}
  \hline\noalign{\smallskip} 
\multicolumn{1}{c}{Parameter} & \multicolumn{1}{c}{Input Value} & \multicolumn{1}{c}{Priors} & \multicolumn{1}{c}{Starting position} & \multicolumn{1}{c}{Fitted value} & \multicolumn{1}{c}{Residual} \\
  \hline\noalign{\smallskip}
$\lambda_{\text{star}}$ & 6613.825 & [6613.0, 6615.0] & 6613.600 & $6613.841^{+0.024}_{-0.025}$ & 0.0156 \\
$\lambda_{\text{DIB}}$  & 6613.709 & [6612.8, 6614.6] & 6613.730 & $6613.714^{+0.006}_{-0.006}$ & 0.0047 \\
$A_{\text{star}}$         & 0.011    & [0.0010, 0.4000]   & 0.010    & $0.0104^{+0.001}_{-0.001}$     & -0.0006 \\
$A_{\text{DIB}}$          & 0.049    & [0.0066, 0.1248]   & 0.046    & $0.0486^{+0.001}_{-0.001}$     & -0.0004 \\
$\sigma_{\text{star}}$  & 0.380    & [0.0014, 1.0000]   & 0.500    & $0.404^{+0.026}_{-0.025}$     & 0.0236 \\
$\sigma_{\text{DIB}}$   & 0.546    & [0.3022, 0.7926]   & 0.537    & $0.549^{+0.006}_{-0.006}$     & 0.0031 \\
  \hline\noalign{\smallskip}
\end{tabular}
\end{center}
\end{table*}

Uniform distributions were adopted as the priors for all six parameters. The prior ranges for the DIB parameters are informed by statistical template matching measurements of 31,244 sources from LAMOST MRS spectra, constrained to a distance within 600 pc based on Gaia EDR3 parallax data (\citealt{2025A&A...702A..58S}). The final boundaries for the DIB parameters are determined by the extremes of this sample after excluding 3$\sigma$ outliers. The boundaries for the stellar line parameters are then appropriately defined relative to the established DIB priors (see Table~\ref{tab:full_fit_results}).

According to Equation~\eqref{eq:dataP}, the logarithmic posterior probability distribution is given by:

\begin{equation}
\ln P(\theta|\text{Data}) \propto \ln L(\theta) + \ln P(\theta)
\end{equation}

The posterior distribution, $P(\theta|\text{Data})$, was sampled using the Markov Chain Monte Carlo (MCMC) method, specifically the Affine Invariant Ensemble sampler implemented in \texttt{emcee} (\citealt{2013PASP..125..306F}). The MCMC method enabled numerical sampling from complex posterior probability distributions. In this six-dimensional parameter space, the algorithm initialized 48 walkers within a Gaussian ball centered around their starting positions, which were set to the median values derived from template matching measurements database (\citealt{2025A&A...702A..58S}). The chains were run for a burn-in phase of 500 steps to ensure convergence, followed by a production run of 5,000 steps to effectively sample the posterior.

\begin{figure*}[!ht]
    \centering
    \includegraphics[ width=0.99\textwidth]{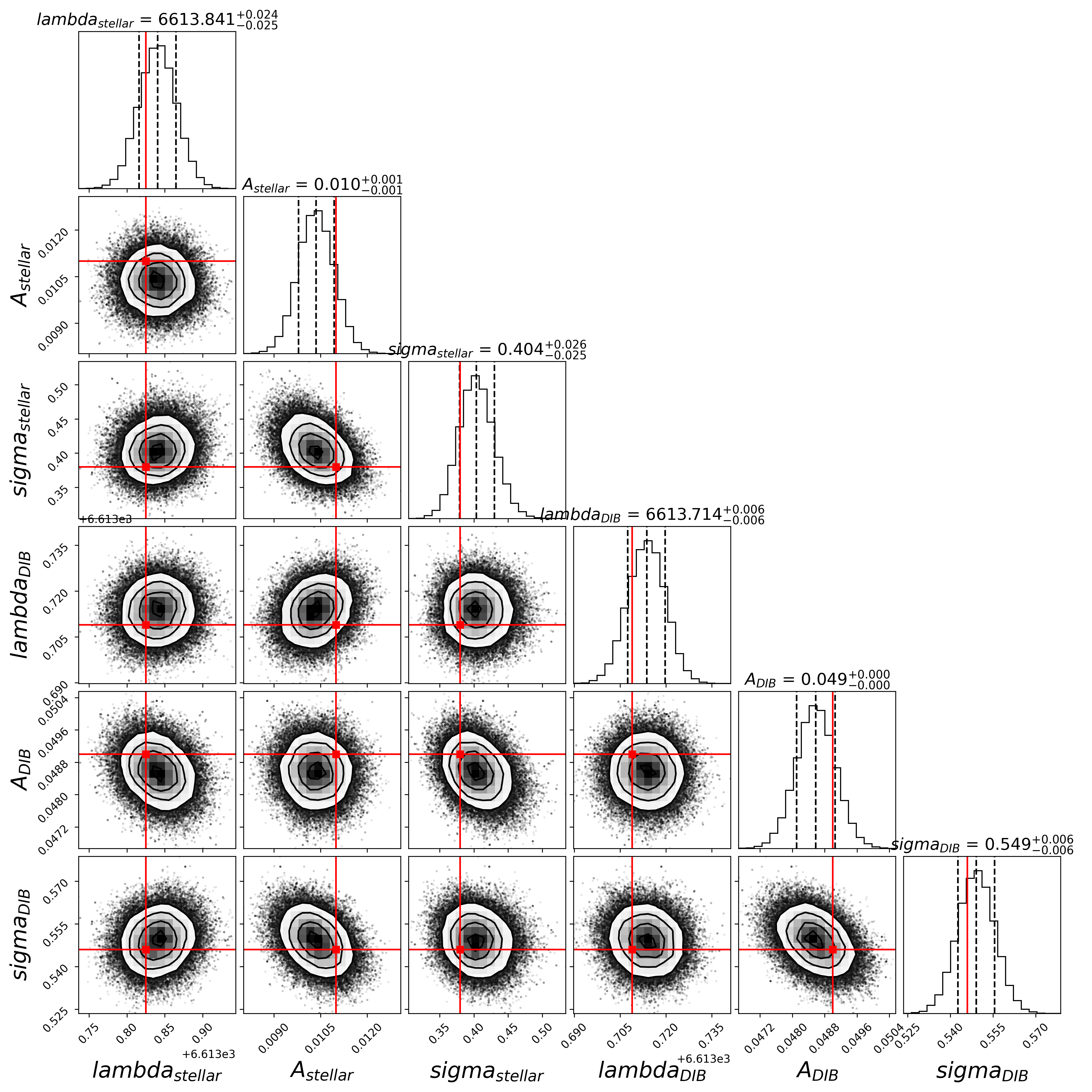} 
    \caption{\label{fig:mcmc_mock}\small Corner plot showing the posterior probability distributions of the fitted parameters for a mock spectrum. The diagonal panels display 1D histograms where all six parameters exhibit Gaussian-like distributions, while the off-diagonal panels show 2D joint distributions. The nearly circular shapes of these joint distributions indicate effective parameter decoupling. The red lines indicate the true values used in the simulation; the intersections of these lines fall within the $3\sigma$ confidence region of the posteriors. The slight offset between the posterior centers and the true values is attributed to systematic bias introduced by RV errors. }
\end{figure*}

\subsection{Posterior Distributions and Algorithm Benchmark Performance}
\label{subsec:Posterior}

We first assessed the algorithm's baseline performance using the mock spectra described in Section~\ref{sec:maths}. Processing 24 spectra (using 48 walkers, 500 burn-in, and 5,000 production steps) requires 73 seconds on an Intel i7-14650HX (16 GB RAM) workstation. 

As illustrated in the corner plot (see Figure~\ref{fig:mcmc_mock}), the diagonal panels display the 1D posterior distributions for each parameter. These distributions were Gaussian-like, showing no significant long tails or multimodal behaviors. The off-diagonal panels show the 2D joint posteriors, which are nearly circular. Notably, the joint distribution of the two spectral lines exhibits a distinct circular shape, indicating that the algorithm effectively disentangled the two lines. The red lines represent the ground truth of the mock spectra; all true values fall well within the $3\sigma$ credible intervals. The slight offset between the true values and the centers of the joint distributions is attributed to the systematic error introduced by the RV error propagation.

Table~\ref{tab:full_fit_results} summarizes the fitting results. The uncertainties are derived from the 16th and 84th percentiles ($1\sigma$ credible intervals) of the posterior distributions. The fitted values agreed closely with the inputs: wavelength offsets are $<0.02\,\rm{\AA}$ (relative deviations $<0.03\%$), depth residuals are $\le 6\times10^{-4}$ ($\sim 5\%$ for $A_{\rm star}$ and $\sim 1\%$ for $A_{\rm DIB}$), and width residuals are $\le 0.024$ ($\sim 6\%$ for $\sigma_{\rm star}$ and $\sim 1\%$ for $\sigma_{\rm DIB}$). This indicates that, under ideal conditions, the algorithm recovers intrinsic DIB profiles with high fidelity.

\section{Validation with Mock Spectra}
\label{sec:Validation}
In Section~\ref{subsec:Posterior}, we demonstrated the feasibility of the MCMC-RVSD algorithm under ideal observational conditions. In this section, We first focused on three key observational parameters (spectral resolution, SNR, and RV measurements), to provide reliable parameter guidance for real observations. We then assessed the generality of the method by relaxing the assumptions underlying the mock spectra, including contamination by multiple stellar lines, a non-Gaussian DIB profile, and the applicability to other DIBs.

In all the tests of this section, the mock spectra are generated with the configuration of Section~\ref{sec:maths}, and only the parameters required by the individual test are varied. The
fiducial profile adopted in Section~\ref{sec:maths} represents the typical DIB $\lambda$6614 measured in the LAMOST MRS ($A_{\text{DIB}}=0.049$, $\sigma_{\text{DIB}}=0.546\,\rm{\AA}$; \citealt{2025A&A...702A..58S}), for which the Gaussian-fit equivalent width $EW_{DIB}=A_{\text{DIB}}\,\sigma_{\text{DIB}}\sqrt{2\pi}\approx0.067\,\rm{\AA}$ corresponds to $E(B-V)\approx0.3$ mag through the empirical $EW_{DIB}$ -- $E(B-V)$ relations (\citealt{2015MNRAS.452.3629L}; \citealt{2011ApJ...727...33F}).

In the simulations, the resulting spectra varied due to noise and random processes, leading to fluctuations in the overall trends of the result. To mitigate fluctuations caused by randomness, three observational parameter settings were computed 30 times repeatedly. The median value was selected to represent the typical performance under a given parameter configuration, and the standard deviation of the probability distribution of the fitted values from the repeated measurements was used as the error bar.

\subsection{Effect of Spectral Resolution}
\label{subsec:Resolution}
To investigate the applicability of the MCMC-RVSD method under low-resolution conditions, tests were conducted at resolutions ranging from R = 500 to 7500 in steps of 500. Since the spectral parameters used to generate the mock spectra are sourced from the LAMOST medium-resolution database, it was not feasible to generate spectra with resolutions exceeding 7500. The method for generating the mock spectra was nearly identical to that described in Section~\ref{sec:maths}. The only difference was that we degraded the spectral resolution before adding noise.

Prior to resolution degradation, the spectrum was interpolated from a sampling step of 0.12 \rm{\AA} to 0.01 \rm{\AA}. This allows the subsequent convolution to employ a finer Gaussian kernel, enabling a clearer distinction between the effects of different resolutions. The convolution of two Gaussian functions results in another Gaussian function. The Full Width at Half Maximum ($FWHM_*$) of the Gaussian kernel used for convolution satisfies

\begin{equation}
FWHM_{*}=\sqrt{FWHM_{final}^2-FWHM_{initial}^2}
\label{eq:FWHM}
\end{equation}

Based on the definition of resolution, $R={\lambda}/{FWHM}$, the required $FWHM_*$ of the convolution kernel can be derived:

\begin{equation}
FWHM_{*}=\lambda \sqrt{\frac{1}{R_{\text {final}}^2}-{\frac{1}{R_{\text {initial}}^2}}}
\label{eq:FWHM-R}
\end{equation}

Given the mock spectrum's wavelength range of 6400–6800 \rm{\AA}, a central wavelength ($\lambda$) of 6600 \rm{\AA} was adopted for simplification. The width of the Gaussian kernel ($\sigma$), was then derived from the definition of FWHM ($2\sqrt{2ln2}\sigma$). A kernel size covering $5\sigma$ (encompassing 99.99\% of the energy) was chosen, divided by the sampling step, and rounded to the nearest odd integer to obtain the size in pixel units. Finally, the convolution method from the ``astropy'' package was used to create the convolution kernel and convolve the input spectrum.

\begin{figure*}[ht]
    \centering
    \includegraphics[width=0.999\textwidth]{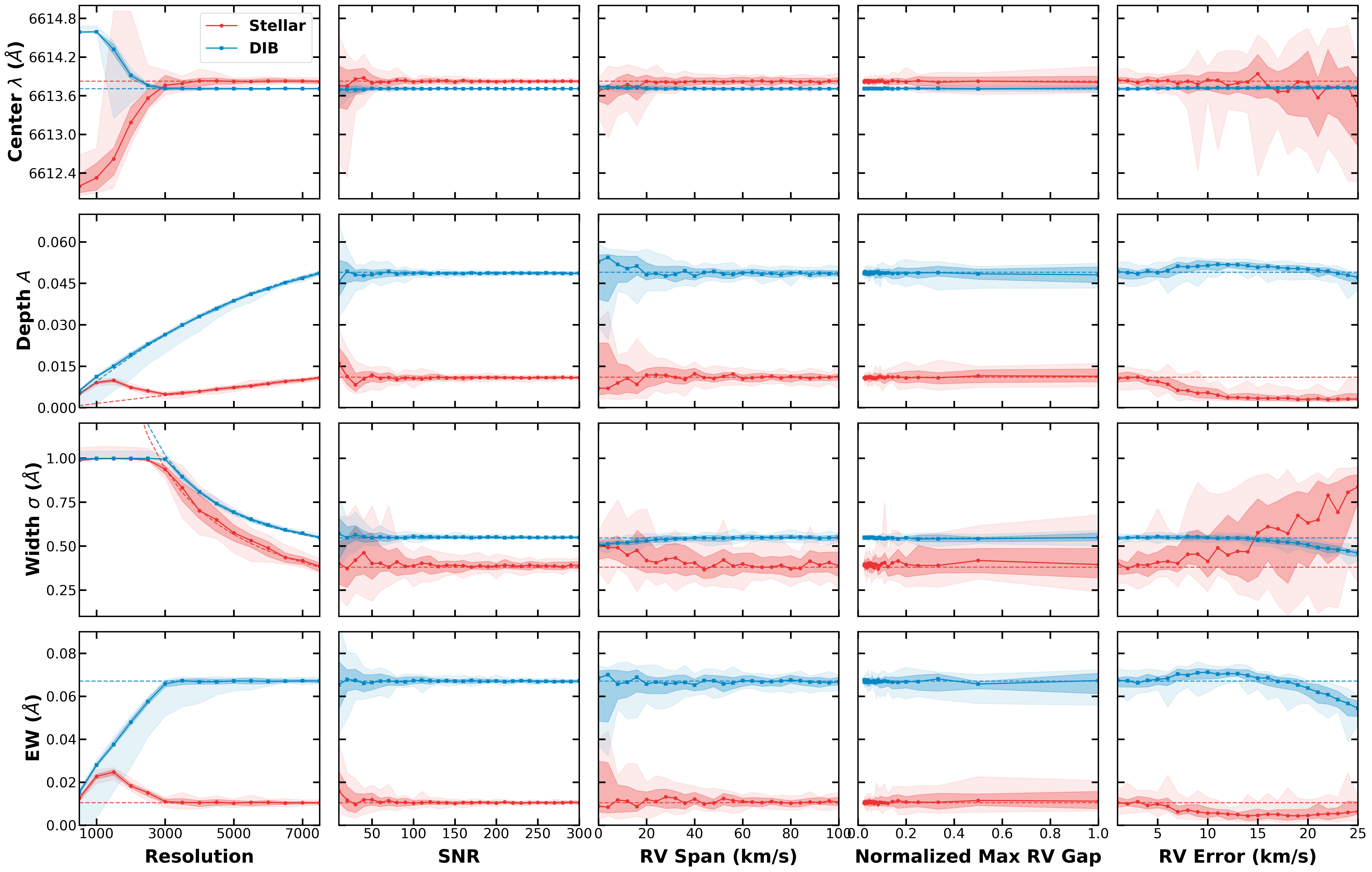}

    \caption{\label{fig:parameter} \small Algorithm performance under varying observational conditions. Each column presents the measured line parameters (central wavelength $\lambda_{\text{DIB}}$, absorption depth $A_{\text{DIB}}$, Gaussian width $\sigma_{\text{DIB}}$, and equivalent width $EW_{\text{DIB}}$) as a function of a observational condition: spectral resolution, SNR, RV span, normalized max RV gap and RV error. The solid lines connect the median values of 30 realizations, with the shaded regions representing the $1\sigma$ and $3\sigma$ scatter, quantifying the algorithm's accuracy and precision.}
\end{figure*}

First column of Figure~\ref{fig:parameter} presents the spectral line parameters measured by the algorithm at different resolutions. As resolution decreases, the line absorption depth decreases and the lines broaden. When the resolution drops below 3000, the extended line wings were truncated due to the spectral cropping and the boundaries of the prior distributions. This truncation causes the measured line width to remain constant, leading to an underestimation of the  EW and a deviation in the measured central wavelength. Consequently, the method presented in this study is not suitable for spectra with resolutions below 3000. This specifically implies that for the  LAMOST low resolution spectra ($\text{R} \approx 1800$), the parameter priors would need to be redesigned.

The tests above assume a single resolution across all epochs of a target. In practice, the resolution varies from fiber to fiber, because the LAMOST fibers have slightly different instrumental profiles (IPs) (\citealt{2026AJ....171..256L}). Using the public LRS IP model of \cite{2026AJ....171..256L}, we measured a fiber-to-fiber FWHM dispersion of $\approx4\%$ ($1\sigma$) at the DIB wavelengths and conservatively assume $\approx5\%$ for the MRS. Repeating the test with non-uniform resolutions (fixed and mixed $R$ between 6500 and 7500, 30 repeats each), the three DIB parameters are recovered within $3\sigma$ in 100\% of the cases and all deviations are $<2\sigma$, with only a shift of the width and depth toward the mean resolution. In extreme cases, such as uneven RV sampling and large resolution differences, the systematic offset may become appreciably larger and the fits may fail to converge. The autocorrelation time criterion in Section~\ref{subsec:label} can identify and exclude such unreliable fits.

\subsection{Effect of Signal-to-Noise Ratio}
\label{subsec:SNR}
The SNR affects both the noise level and flux error in spectra, see Equations~\eqref{eq:chi-ferr}--\eqref{eq:snr-chi}. To evaluate the robustness of the MCMC-RVSD method under different noise conditions, mock spectra were generated by varying only the SNR parameter.

The results in the second column of Figure~\ref{fig:parameter} show that for LAMOST MRS data, the errors of the fitted line parameters increase sharply when the SNR falls below 100. It is also observed that stellar line parameters were fitted with lower accuracy and stability compared to the DIB features. This is because stellar lines have shallower absorption depths and lower equivalent widths, making them more susceptible to noise. For DIBs of different strengths, the impact of low SNR diminishes for stronger lines. Furthermore, our algorithm incorporates an SNR-based weighting scheme in the likelihood function, see Equation~\eqref{eq:likelihood snr}, which reduces the influence of spectra with poor SNR on the fit. It is preliminarily recommended to use spectra with $\text{SNR} > 50$. A more detailed analysis and SNR recommendations based on real observational data will be provided in Section~\ref{subsec:correlation}.

\subsection{Effect of Radial Velocity Parameters}
\label{subsec:Radial Velocity}

RV is the core parameter for distinguishing the moving stellar component from the stationary DIB feature. To assess the MCMC-RVSD algorithm's sensitivity to the quality of RV information, we systematically tested the effects of three key RV-related observational parameters: the RV sampling span (RV span), the normalized maximum RV sampling gap (RV gap), and the RV error. Mock spectra were generated by varying these parameters individually, following the baseline configuration in Section~\ref{sec:maths}.

The RV span, defined as the total range of Doppler shifts sampled (e.g., from $-20$ to $+20$ km s$^{-1}$ corresponds to an RV span of $40$ km s$^{-1}$). In our tests, an array of 5 uniformly spaced RVs (e.g., $[-2, -1, 0, 1, 2]$ km s$^{-1}$) was scaled to achieve different spans, with a fixed RV error of $1.36$ km s$^{-1}$ (representative of LAMOST MRS data). As shown in the third column of Figure~\ref{fig:parameter}, the fitting errors were largest when the RV span is zero (no differential RV). The errors decrease significantly as the span increases, confirming that the algorithm successfully leveraged RV separation to constrain the line parameters. A span of less than $20$ km s$^{-1}$ results in relatively large errors; beyond this point, the precision gain plateaus as other factors become dominant.

The RV gap, defined as the largest interval between consecutive RVs normalized by the total span, quantifies the uniformity of orbital phase sampling (\citealt{2024ExA....57...12K}). Tests were conducted with a fixed RV span of $40$ km s$^{-1}$ while increasing the number of uniformly spaced sampling points. A smaller gap indicates more uniform sampling (e.g., a gap of 1 for 2 points, 0.5 for 3 points). The fourth column of Figure~\ref{fig:parameter} shows that the impact of the RV gap on final parameter errors only became noticeable when the gap is smaller than approximately 0.2 (i.e., with more than 5-6 well-distributed epochs).

\begin{figure*}[!htp]
  \centering
  \includegraphics[width=0.99\textwidth]{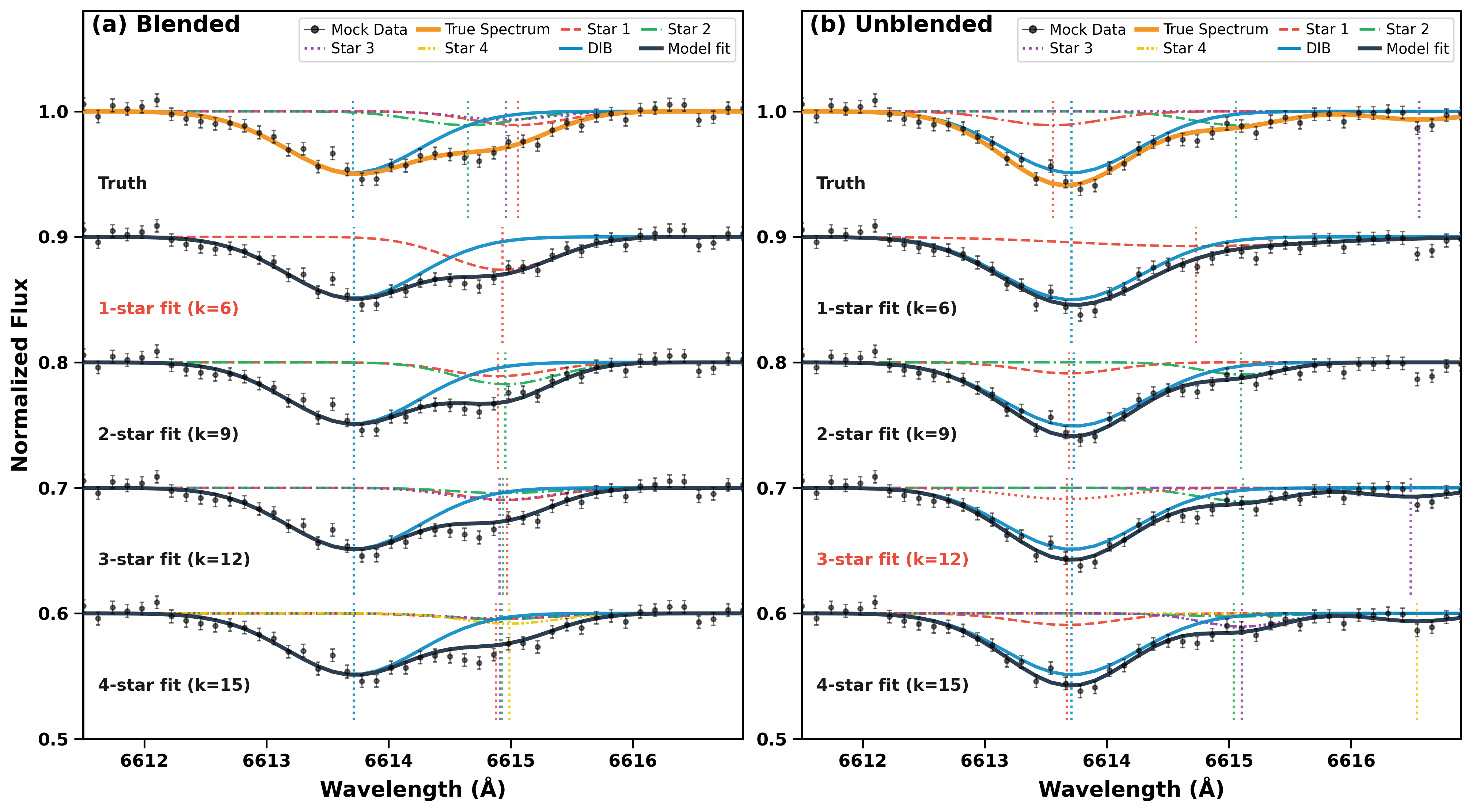}
  \caption{\label{fig:multi_spec}{\small MCMC-RVSD iterative fitting of three stellar lines, for both the blended and the unblended case. The left (blended) panel adopts real stellar-line central wavelengths from the NIST atomic database, and a 1-star model already provides a good fit. The right (unblended) panel shows mock data with unblended stellar lines, and the 3-star model successfully recovers the three stellar components.}}
\end{figure*}

The RV error represents the measurement uncertainty at each epoch. We tested a range of errors from $0$ to 25 ~km~s$^{-1}$ under otherwise identical conditions. The last column of Figure~\ref{fig:parameter} indicates that RV errors below $5 \mathrm{~km~s^{-1}}$ had a negligible effect on the fitting quality. However, for errors exceeding $10 \mathrm{~km~s^{-1}}$, the uncertainties of the stellar line parameters (which are weaker) increased markedly, and their fitted absorption depths and equivalent widths begin to bias toward zero. This suggests that large RV errors can cause the algorithm to effectively ``ignore'' the weaker stellar component. In some cases, the true parameter values fall outside the $3\sigma$ posterior credible intervals. This breakdown was partly attributable to the first-order Taylor linear approximation used in the likelihood function to propagate RV errors (Section~\ref{subsec:Likelihood}), whose accuracy becomes inadequate for RV error $> 10\mathrm{~km~s^{-1}}$. In addition, for LAMOST MRS, almost all spectra with $\text{SNR}>50$ have RV errors below $5 \mathrm{~km~s^{-1}}$. For data from other telescopes and other RV measurement methods, however, the RV errors may exceed this boundary. Moreover, for other strong ISM line, the SNR $> 50$ threshold is not mandatory. Excluding sources with RV error $> 5 \mathrm{~km~s^{-1}}$ is necessary when selecting the sample.

\subsection{Multiple Stellar Lines}
\label{subsec:MultiStar}

To handle the case where more than one stellar line contaminates a DIB feature, we adopted a two-step approach. First, in the spectral-fitting stage, we masked out the region beyond $3\sigma$ of the DIB center. Second, we iteratively added stellar line models with central wavelengths taken from the NIST atomic database (\citealt{NIST_ASD}), progressively increasing the number of stellar lines $N$ according to line strength, and selected the optimal $N$-star model using the (Akaike Information Criterion) and BIC (Bayesian Information Criterion). When the AIC or BIC minus its minimum equals 0 ($\Delta$AIC or $\Delta$BIC = 0), the current model strikes the optimal balance between underfitting and overfitting.

\begin{figure}
  \centering
  \includegraphics[width=0.99\linewidth]{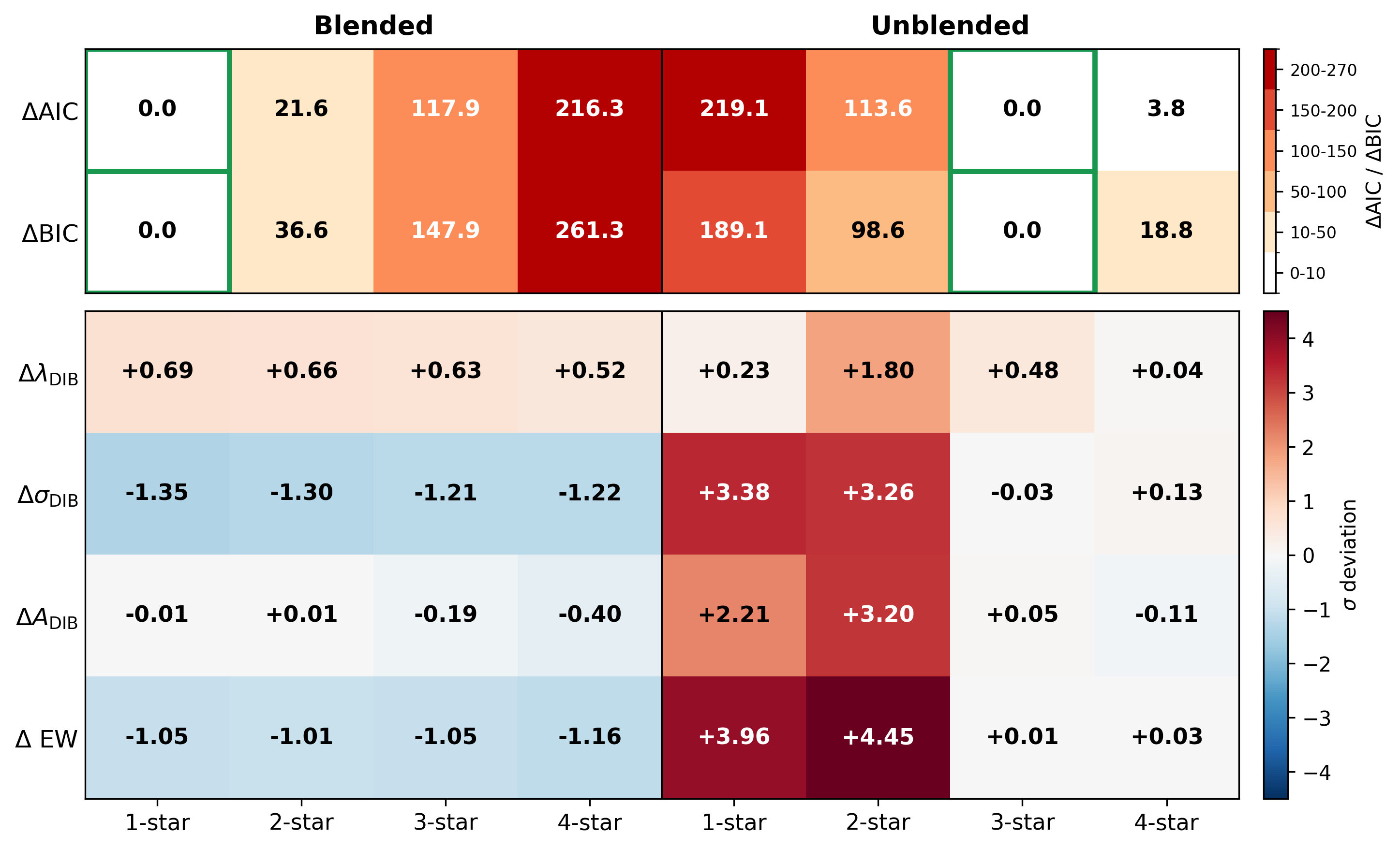}
  \caption{\label{fig:multi}{\small Model-selection criteria and DIB parameter residuals for the blended (left) and unblended (right) stellar-line configurations. The top two rows show the $\Delta$AIC and $\Delta$BIC values relative to their minima, rendered with a discrete color scale in which the white cell marks the optimal model (outlined in green). The bottom four rows show the residuals of the four DIB parameters in units of the $1\sigma$ uncertainty. In the blended configuration the three stellar lines are unresolved and the 1-star model minimizes both AIC and BIC, whereas in the unblended configuration the 3-star model is optimal; the 1-star and 2-star models then overestimate the DIB strength by more than $3\sigma$.}}
\end{figure}

To test the impact of multiple stellar lines on DIB measurements, we tested this with two mock configurations. In configuration (a) of Figure~\ref{fig:multi_spec}, the three real stellar lines near DIB $\lambda\,6613.709$~\AA---Fe~I ($6613.825$~\AA), Th~I ($6613.373$~\AA), and Y~II ($6613.731$~\AA)---are severely blended. Their profile parameters come from the fit result of solar analogs using SpectroWeb 2.0 database (\citealt{1966sst..book.....M,2011CaJPh..89..395L}). $\Delta$AIC and $\Delta$BIC in Figure~\ref{fig:multi} select the 1-star model. Inspecting the fit quality in more detail, the deviations of the line parameters measured by the 1-star model from the true values are almost at the same level as those of the other models. The reason is that LAMOST MRS resolution of $R=7500$ corresponds to an instrumental FWHM of $\lambda/R\approx0.88$~\AA{} at 6614~\AA. These three stellar lines are so severely blended that they can be treated as a single stellar line.

To better demonstrate the advantage of the iterative fitting, in configuration (b) we shifted the central wavelengths of the three stellar lines to simulate stellar lines that are well separated from each other. They are set to $\lambda\,6613.825\,\AA$, $\lambda\,6612.325\,\AA$, and $\lambda\,6615.325\,\AA$. Configuration (b) of Figure~\ref{fig:multi_spec} shows that the three stellar lines are clearly unblended and $\Delta$AIC and $\Delta$BIC in results indicate that the 3-star model is optimal. Compared with the true values in Figure~\ref{fig:multi}, the 1-star and 2-star models overestimate the strength of the DIB, enlarging the width and depth of the DIB line. The 4-star model shows deviations consistent with the 3-star model, but its $\Delta$AIC and $\Delta$BIC are slightly larger.

For multiple stellar line contamination, we iteratively add stellar components to the MCMC-RVSD and identify severe contamination. For faint or blended stellar lines, the spectrum provides insufficient information to remove them, yet they do not significantly affect the DIB extraction. For unblended lines, we selected the model minimizing AIC or BIC, following Occam's razor. It is worth noting that, the AIC and BIC can exhibit multiple local minima, owing to the large number of parameters and the non-smoothness of the MCMC fit. We adopted the first local minimum and cap the number of stellar lines according to the atomic lines provided by NIST, so that a minimum is always attained.

\subsection{Non-Gaussian DIB Profile}
\label{subsec:nongauss}

In the Bayesian framework of Section~\ref{sec:Bayesian}, we modelled the profile of the DIB $\lambda$6614 as a Gaussian. For DIB $\lambda$6614, a Gaussian profile does recover the equivalent width reliably in LAMOST MRS data (\citealt{2025A&A...702A..58S}). For broad or very high resolution DIBs, however, the profiles are manifestly non-Gaussian: at $R=220{,}000$, DIB $\lambda$6614 splits into three principal components (\citealt{1996MNRAS.283L.105K,2002A&A...384..215G,2004ApJ...611L.113C}), the weak DIB $\lambda$4964 has an asymmetric profile with an extended wing towards longer wavelengths (\citealt{2002A&A...396..987G}), and DIB $\lambda$4428 is close to a Lorentzian shape (\citealt{2002ApJ...578..877S}).

\begin{figure}
    \begin{minipage}[t]{0.995\linewidth} 
    \centering
    \includegraphics[width=1\linewidth]{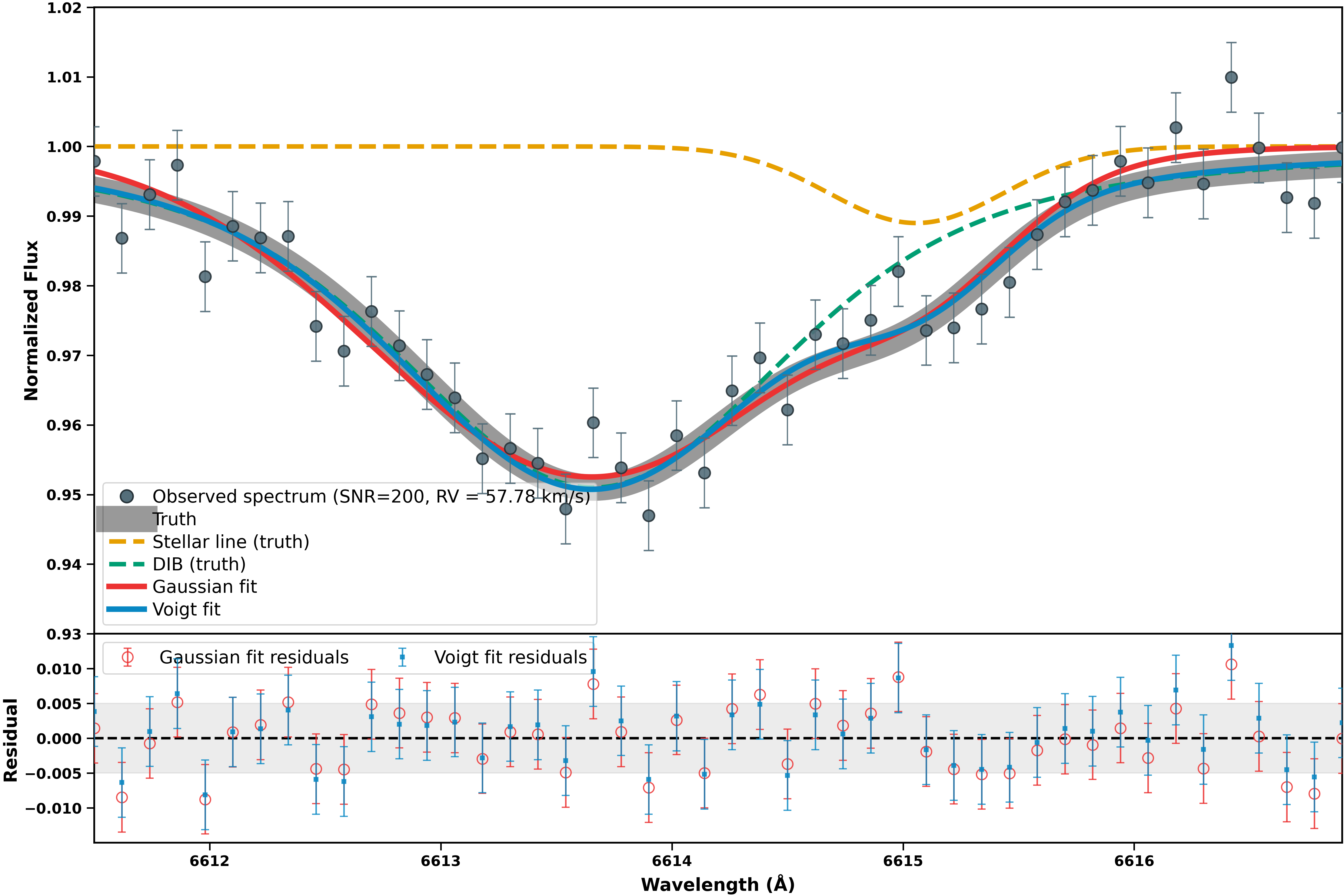} 
    \caption{\label{fig:nongauss}\small The upper panel shows fits to mock spectra (dot) whose intrinsic DIB profile (green dashed line) is a Voigt function
    ($\gamma \approx 0.6\,\rm{\AA}$, i.e. $\tau = 2\,\mathrm{ps}$), with a stellar line (yellow dashed line). The six-parameter Gaussian fit (red) absorbs the Lorentzian wings of DIB $\lambda$6614 into the line width. The seven-parameter Voigt fit (blue) reproduces the line wings and is close to the truth line (black). The lower panel shows the fit residuals, the shaded band
    indicates the noise level.}
    \end{minipage}
\end{figure}

To test the flexibility of our method against non-Gaussian profiles, we generated a set of mock LAMOST MRS spectra whose intrinsic DIB profile is a Voigt function (Figure~\ref{fig:nongauss}), and compared Gaussian fits with fits from a Voigt model. A Voigt profile, the convolution of a Gaussian and a Lorentzian, describes both the homogeneous and the inhomogeneous broadening of a line. For the Gaussian component of the DIB we adopted a typical DIB width ($\sigma = 0.545\,\rm{\AA}$). Lorentzian component $\gamma$ we used the Lorentzian width to lifetime relation(\citealt{2016A&A...590A..52O}):

\begin{equation}
 \mathrm{FWHM}\,[\mathrm{cm}^{-1}] = \frac{1}{2\pi c\,\tau} = \frac{5.3}{\tau\,[\mathrm{ps}]}
\label{eq:lorentz-tau}
\end{equation}

where $\gamma$ is the half-width at half-maximum of the Lorentzian ($\mathrm{FWHM} = 2\gamma$ in the same units).

\begin{table*}
  \caption[]{\small Results of the Gaussian and Voigt fits to a mock spectrum whose intrinsic DIB profile is a Voigt function with $\gamma = 0.6\,\rm{\AA}$. The Gaussian fit departs from the input $\sigma_{\text{DIB}}$ by $+44.4\sigma$ and from the input $EW_{\text{DIB}}$ by $-2.9\sigma$, whereas the Voigt fit recovers the profile parameters well.}
  \label{tab:q3}
  \begin{center}
  \begin{tabular}{c|c|c|r r|r r}
  \hline\noalign{\smallskip}
\multicolumn{1}{c}{Parameter} & \multicolumn{1}{c}{Input Value} & \multicolumn{1}{c}{Priors} & \multicolumn{2}{c}{Gaussian fit} & \multicolumn{2}{c}{Voigt fit} \\
\multicolumn{1}{c}{} & \multicolumn{1}{c}{} & \multicolumn{1}{c}{} & \multicolumn{1}{c}{Fitted value} & \multicolumn{1}{c}{Deviation} & \multicolumn{1}{c}{Fitted value} & \multicolumn{1}{c}{Deviation} \\
  \hline\noalign{\smallskip}
$\lambda_{\text{star}}$ ($\,\rm{\AA}$) & 6613.825 & [6612.0, 6615.0] & $6613.859^{+0.024}_{-0.025}$ & $+1.4\sigma$ & $6613.827^{+0.022}_{-0.022}$ & $+0.1\sigma$ \\
$A_{\text{star}}$ & 0.011 & [0.0010, 0.4000] & $0.010^{+0.001}_{-0.001}$ & $-1.7\sigma$ & $0.011^{+0.001}_{-0.001}$ & $-0.3\sigma$ \\
$\sigma_{\text{star}}$ ($\,\rm{\AA}$) & 0.380 & [0.0014, 1.0000] & $0.378^{+0.029}_{-0.027}$ & $-0.1\sigma$ & $0.358^{+0.024}_{-0.023}$ & $-0.9\sigma$ \\
$EW_{\text{star}}$ ($\,\rm{\AA}$) & 0.010 & --- & $0.009^{+0.001}_{-0.001}$ & $-1.3\sigma$ & $0.010^{+0.001}_{-0.001}$ & $-1.1\sigma$ \\
$\lambda_{\text{DIB}}$ ($\,\rm{\AA}$) & 6613.660 & [6612.8, 6614.6] & $6613.653^{+0.008}_{-0.008}$ & $-0.9\sigma$ & $6613.649^{+0.008}_{-0.008}$ & $-1.4\sigma$ \\
$A_{\text{DIB}}$ & 0.049 & [0.0010, 0.2000] & $0.047^{+0.001}_{-0.001}$ & $-3.8\sigma$ & $0.049^{+0.001}_{-0.001}$ & $+0.6\sigma$ \\
$\sigma_{\text{DIB}}$ ($\,\rm{\AA}$) & 0.545 & [0.3022, 0.7926] & $0.945^{+0.009}_{-0.009}$ & $+44.4\sigma$ & $0.575^{+0.034}_{-0.035}$ & $+0.9\sigma$ \\
$\gamma_{\text{DIB}}$ ($\,\rm{\AA}$) & 0.600 & [0.0010, 1.0000] & --- & --- & $0.553^{+0.041}_{-0.042}$ & $-1.1\sigma$ \\
$EW_{\text{DIB}}$ ($\,\rm{\AA}$) & 0.114 & --- & $0.111^{+0.001}_{-0.001}$ & $-2.9\sigma$ & $0.114^{+0.001}_{-0.001}$ & $-0.2\sigma$ \\
  \hline\noalign{\smallskip}
\end{tabular}
\end{center}
\end{table*}

Broad DIBs ($10$--$100\,\mathrm{cm}^{-1}$), such as $\lambda$4428, can have $\tau \sim 0.06\,\mathrm{ps}$ (\citealt{2002ApJ...578..877S,2016A&A...590A..52O}), whereas narrow DIBs have $\tau > 1\,\mathrm{ps}$ (\citealt{2016A&A...590A..52O}), as is the case for DIB $\lambda$6614. Here we adopted $\tau = 2\,\mathrm{ps}$ -- the excited-state lifetime measured in the laboratory for $C_{60}^{+}$ (\citealt{2015Natur.523..322C}), a candidate carrier of DIBs, corresponding to $\gamma \approx 0.6\,\rm{\AA}$. Stellar lines were again described by Gaussian, and all other settings ($R=7500$, $\text{SNR}=200$, 24 epochs) were kept identical to Section~\ref{sec:maths}. For the same set of mock spectra, we then ran MCMC fits with a six-parameter Gaussian model and with a seven-parameter Voigt model in which $\gamma$ is an additional free parameter.

The EW recovered from the Gaussian fits agree closely with the Voigt input values, with a bias below $3\%$. As shown in Figure~\ref{fig:nongauss}, however, the Gaussian fit absorbs the Lorentzian wings into the line width. As shown in Table~\ref{tab:q3}, Gaussian fit overestimates $\sigma_{\text{DIB}}$ in a systematic way, by $0.40\,\rm{\AA}$ ($+44\sigma$), while the depth is slightly underestimated. With the Voigt model, $\gamma$ is recovered as $0.553 \pm 0.041\,\rm{\AA}$ ($-1.1\sigma$), the $\sigma_{\text{DIB}}$ bias drops to $0.9\sigma$ ($0.575 \pm 0.035\,\rm{\AA}$), and the line wings follow the true profile closely.

In summary, both the Gaussian and the Voigt profile recover the EW of the line, but non-uniform broadening inflates the width inferred from a Gaussian fit and makes its uncertainty unreliable. For broader DIBs the carrier lifetime is shorter and the Lorentzian component more prominent, so that a Gaussian fit becomes even broader once the wings are absorbed. Extending MCMC-RVSD to broader, asymmetric, or multi-component DIBs will thus require dedicated models of the profile shape.

\subsection{More DIBs}
\label{subsec:Moredib}

All the validation presented so far concerns DIB $\lambda$6614 alone, a relatively strong ($A_{\text{DIB}} \approx 0.049$), typical width ($\sigma_{\text{DIB}} \approx 0.55\,\rm{\AA}$) and $E(B-V)\approx0.3$ mag line. The applicability of the method elsewhere in the DIB parameter space still needs to be tested. On LAMOST MRS, the DIBs already reported in the literature are $\lambda$6379, $\lambda$6614 and $\lambda$6660 (\citealt{2022RAA....22h5007W}).

To test the generality of the method for other DIBs, we kept the stellar line and all other settings unchanged ($R=7500$, $\text{SNR}=200$, 24 epochs) and varied only the width and depth of the mock DIB: $\sigma_{\text{DIB}}$ from 0.4 to $0.8\,\rm{\AA}$ in steps of $0.1\,\rm{\AA}$, and $A_{\text{DIB}}$ from $0.01$ to $0.10$ in steps of $0.01$, each setting decomposed with the same MCMC settings as in Section~\ref{sec:maths}.

\begin{figure*}
    \centering
    \includegraphics[width=0.999\textwidth]{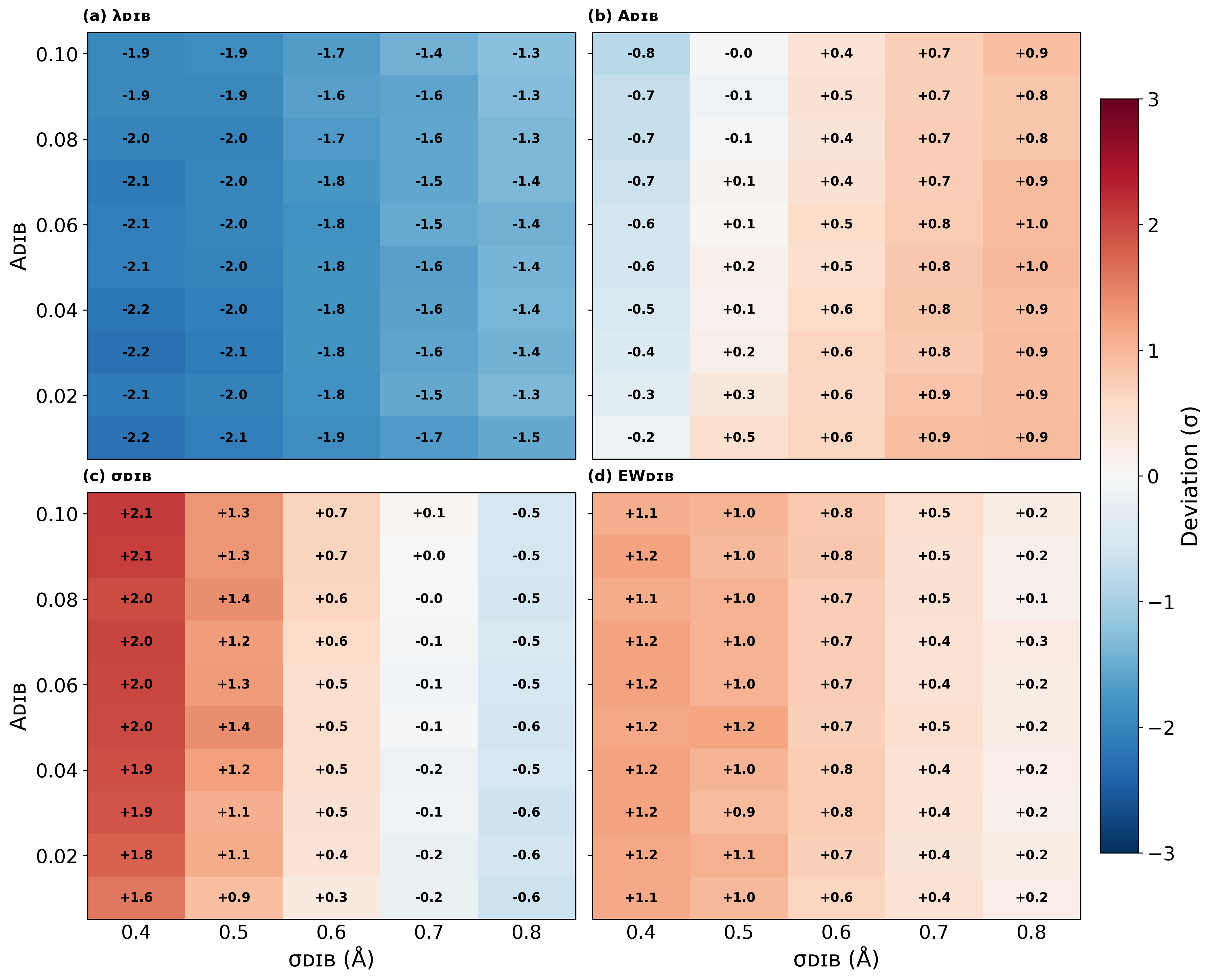}
    \caption{\label{fig:moredib} \small Deviations of the MCMC-RVSD recovery across the DIB parameter scan. Each panel shows, for 50 mock spectra, the fitted-to-input offset of (a) $\lambda_{\text{DIB}}$, (b) $A_{\text{DIB}}$, (c) $\sigma_{\text{DIB}}$ and (d) $\text{EW}_{\text{DIB}}$, in units of the posterior $1\sigma$ error, against the input $\sigma_{\text{DIB}}$ and $A_{\text{DIB}}$. All 50 combinations are recovered within $3\sigma$ (maximum $2.23\sigma$), the deviations are set by the line width and are almost independent of the DIB depth.}
\end{figure*}

As shown in Figure~\ref{fig:moredib}, the parameter deviations remain below $3\sigma$ in all 50 cases (maximum $2.23\sigma$; 15 cases exceed $2\sigma$), and they depend mainly on the width, not the depth.  For the narrowest DIBs ($\sigma_{\text{DIB}} = 0.4\,\text{\AA}$, covering only about three sampled pixels) the deviations are largest. The line width not only leads to an overestimate of $\sigma_{\text{DIB}}$ and $\text{EW}_{\text{DIB}}$, but also slightly biases the line-centre measurement. As the width increases the algorithm resolves the DIB component better, and the parameter deviations become smaller. The depth has no significant effect on $\text{EW}_{\text{DIB}}$, but in narrow DIBs it increases the fitted line width. In addition, since all tests use the same set of RV errors, a systematic offset in the central wavelength parameters is expected.

It should be noted that varying the width and depth of a single DIB does not cover all DIB cases. In this test, DIBs narrower than $\sigma_{\text{DIB}} \lesssim 0.35\,\rm{\AA}$ (such as $\lambda$6660, $\sigma \approx 0.30\,\rm{\AA}$) and broader than $\sigma_{\text{DIB}} \gtrsim 0.9\,\rm{\AA}$ approach the bounds of the adopted priors. In measurements of real spectra, $\lambda$6379 is likewise readily detected, whereas the much narrower $\lambda$6660 is difficult to measure. For narrower DIBs, the prior should adjusted based on their line profiles. For broad DIBs ($10$--$100\,\mathrm{cm}^{-1}$), besides the effect of the number of local stellar lines (Section~\ref{subsec:MultiStar}), the Lorentzian component and the line wings can no longer be neglected and must be included explicitly in the model (Section~\ref{subsec:nongauss}).

\subsection{Summary}

For DIB $\lambda 6614$ at typical profile ($A_{\text{DIB}}=0.049$,\  $\sigma_{\text{DIB}}=0.546\,\rm{\AA}$,\ $E(B-V)\approx0.3$ mag), extensive testing on mock spectra revealed the key observational constraints for reliable application of the MCMC-RVSD algorithm, summarized as follows:

\begin{itemize}
    \item \textbf{Resolution:} The algorithm is suitable for spectra with resolution $\text{R}>3000$. For $\text{R}<3000$, systematic errors arise from spectral cropping and prior boundary effects, leading to underestimated EW and biased central wavelengths. The method is not applicable to LAMOST low-resolution ($\text{R}\approx 1800$) spectra without revised priors.
    \item \textbf{Signal-to-Noise:} High SNR is essential. For LAMOST MRS data, fitted parameter uncertainties increase sharply when $\text{SNR} < 100$. We recommend $\text{SNR} > 50$ for reliable results, noting that the weaker component is more susceptible to noise.
    \item \textbf{RV Span:} A large total radial velocity span is critical for separating the stellar and DIB components. $\text{RV span} > 20\ \text{km}\ \text{s}^{-1}$ is required to achieve well-constrained parameters with low uncertainties.
    \item \textbf{RV Gap and Error:} Dense, uniform orbital phase sampling (normalized max RV gap $< 0.2$) further reduces fitting errors. The algorithm is robust to $\text{RV error} < 5\ \text{km}\ \text{s}^{-1}$. But performance degrades for $\text{RV error} > 10\ \text{km}\ \text{s}^{-1}$ as true values exceed the $3\sigma$ confidence range, due to the failure of the linear Taylor expansion in the likelihood function.
    \item \textbf{Model Complexity:} Additional stellar lines and other DIBs are recovered within $3\sigma$ once the corresponding components are included in specific models. Contamination by multiple stellar lines can be handled by adding stellar components step by step, with $\Delta\text{AIC}$ and $\Delta\text{BIC}$ identifying the optimal model. For a non-Gaussian DIB profile that contains a Lorentzian component, a Voigt profile should be adopted: a Gaussian profile absorbs the line wings and makes the measured width unreliable, although the effect on $EW_{\text{DIB}}$ remains small. For measurements of other DIBs, the priors of the model must be set independently. And for narrower or weaker DIBs, both the width and the $EW_{\text{DIB}}$ are overestimated. For more complex cases in real measurements, the quality labels of Section~\ref{subsec:label} indicate the reliability of the results.
\end{itemize}

These criteria establish the practical boundaries for applying the method, which is validated to be effective for spectra meeting these conditions, enabling its application to real multi-epoch data.

\section{Application to Observational Data}
\label{sec:realdata}

In this section, we applied the MCMC-RVSD method to real observational data to validate its practicality and robustness. We first constructed a multi-epoch sample by cross-matching the Gaia DR3 non-single star (NSS) catalog (\citealt{2023A&A...674A...1G}) with the LAMOST MRS DIB catalog (\citealt{2025A&A...702A..58S}), and then performed uniform preprocessing and MCMC fitting. We classified the fitting results into three quality tiers---``Success", ``Mild Degeneracy", and ``Severe Degeneracy"---based on the autocorrelation time of the MCMC chains ($\tau_{\rm DIB}$), and identified the key observables (RV span, RV error, SNR, and others) governing the fitting quality. Finally, we presented a labeled catalog of DIB $\lambda 6614$ measurements for the 434 binary systems.

\subsection{Observational Data}
To evaluate our MCMC-RVSD algorithm against the traditional template matching method in a realistic observational context, we cross-matched a sample of 31,244 sources from an existing template-measured DIB catalog with the Gaia DR3 ``nss\_two\_body\_orbit" catalog. This initial cross-match yielded 992 binary systems, predominantly consisting of ``Orbital" and ``SB1" types. We subsequently cross-matched these targets with the LAMOST DR13 v1.0 Medium-Resolution Multiple-Epoch Catalog. To ensure reliability, we filtered out low-quality spectra by imposing a SNR threshold of 50 and requiring at least two observational epochs per source. Finally, we obtained a final sample of 434 valid sources with 1,529 individual spectra.

Before applying the MCMC-RVSD algorithm, we processed the raw LAMOST spectra to ensure high data quality. The preprocessing steps were structured as follows:

First, we applied pixel-level masking using the \texttt{andmask} and \texttt{ormask} flags provided by the LAMOST pipeline for eliminating invalid pixels affected by cosmic rays or detector saturation. We converted the wavelength solution from vacuum wavelengths to air wavelengths to ensure alignment with standard atomic line databases. Next, we trimmed the spectra strictly to the DIB localized feature range ($6604$--$6664\,\rm{\AA}$) and linearly resampled both the flux and inverse variance arrays onto a uniform wavelength grid with a constant step size of $0.12\,\rm{\AA}$.

We adopted a 4th-order polynomial to fit the spectrum, coupled with an asymmetric sigma-clipping algorithm that heavily penalized and rejected structured absorption features (i.e., iteratively excluding localized pixels falling $< 1\sigma$ below or $> 3\sigma$ above the fit). This iterative process was repeated 20 times to map and robustly establish a local pseudo-continuum.

Finally, we applied a dynamic mask by setting the flux values of regions outside the core analysis window ($6611.5$--$6617\,\rm{\AA}$) to exactly unity (1.0). This spatial boundary condition step crucially prevents irrelevant surrounding spectral structures or wide-scale extended noises from injecting biases into the likelihood function calculations.

\subsection{Quality Labels}
\label{subsec:label}
As fitting quality depends on multiple factors, it was difficult to determine the fit quality based solely on observational parameters. Therefore, we adopted the autocorrelation time ($\tau$) of the MCMC chains as a reference. The $\tau$ represents the number of steps required to produce a statistically independent sample. According to the author of the \texttt{emcee} package (\citealt{2013PASP..125..306F}), when the chain length ($L$) exceeds 50 times the $\tau$, the samples were considered sufficient and independent enough to describe the target probability distribution.

Therefore, we classified the fitting results based on their $\tau$. In our measurement window, stellar lines were prone to blending and interference from other stars in the binary system. In contrast, the stationary interstellar lines rarely blend together, and they are our primary targets. Thus, we used the $\tau$ of the three DIB parameters as the classification thresholds, denoted as $\tau_{\text{DIB}}$, including $\tau({\lambda _{\text{DIB}}})$, $\tau({A _{\text{DIB}}})$ and $\tau({\sigma _{\text{DIB}}})$.

\begin{figure*}[ht]
    \centering
    \includegraphics[ width=0.8\textwidth]{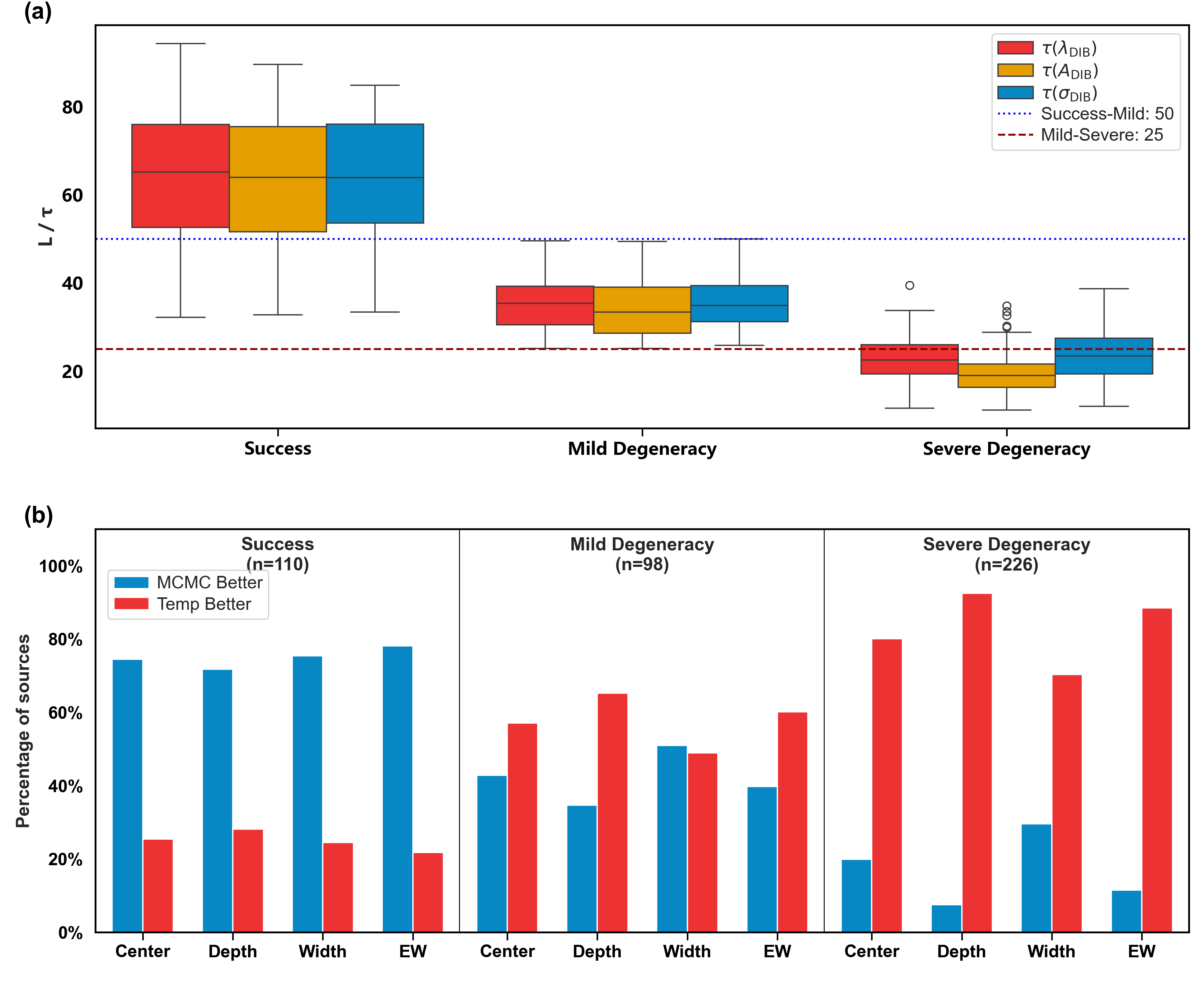} 
    \caption{\label{fig:combine} \small (a)Distribution of the autocorrelation times ($\tau_{\mathrm{DIB}}$) for all analyzed observation paths. The horizontal boxplot limits demarcate the classification criteria zones. (b)Comparison of the measurement errors (e.g., central wavelength or equivalent width) derived from the MCMC-RVSD method and the traditional template matching method. The performance is grouped according to the $\tau$-based classification, showing a clear correlation between fit quality (autocorrelation time) and consistency with an independent method.}
\end{figure*}

Since a single threshold cannot perfectly distinguish the fit quality, we applied the following rules. For high-quality fits, we relaxed the condition to include more sources: a source is classified as ``Success" if $max(L/\tau_{\mathrm{DIB}} )\ge 50$. For sources with very poor convergence, we strictly enforced the lower limit: a source is classified as ``Severe Degeneracy" if $min(L/\tau_{\mathrm{DIB}} )\le 25$. Sources that fall into neither the ``Success" nor ``Severe" class are classified as ``Mild Degeneracy". Figure \ref{fig:combine} (a) showed that the $\tau$ classification effectively distinguishes the three types, with significant differences in their medians. And we compared the parameter errors with the template matching method in different class. Figure \ref{fig:combine} (b) shows the ``Success" class achieved 80\% sources in getting smaller errors, while the ``Severe" class reached only 10\%. A smaller $\tau$ indicates faster convergence of the MCMC chains, highlighting the algorithm's advantage in high precision. In contrast, a large $\tau$ suggests that the combined factors (RV, SNR, among others) lack sufficient decoupling power, causing the performance to lag behind template matching.

\begin{figure*}
    \centering
    \includegraphics[width=0.8\textwidth]{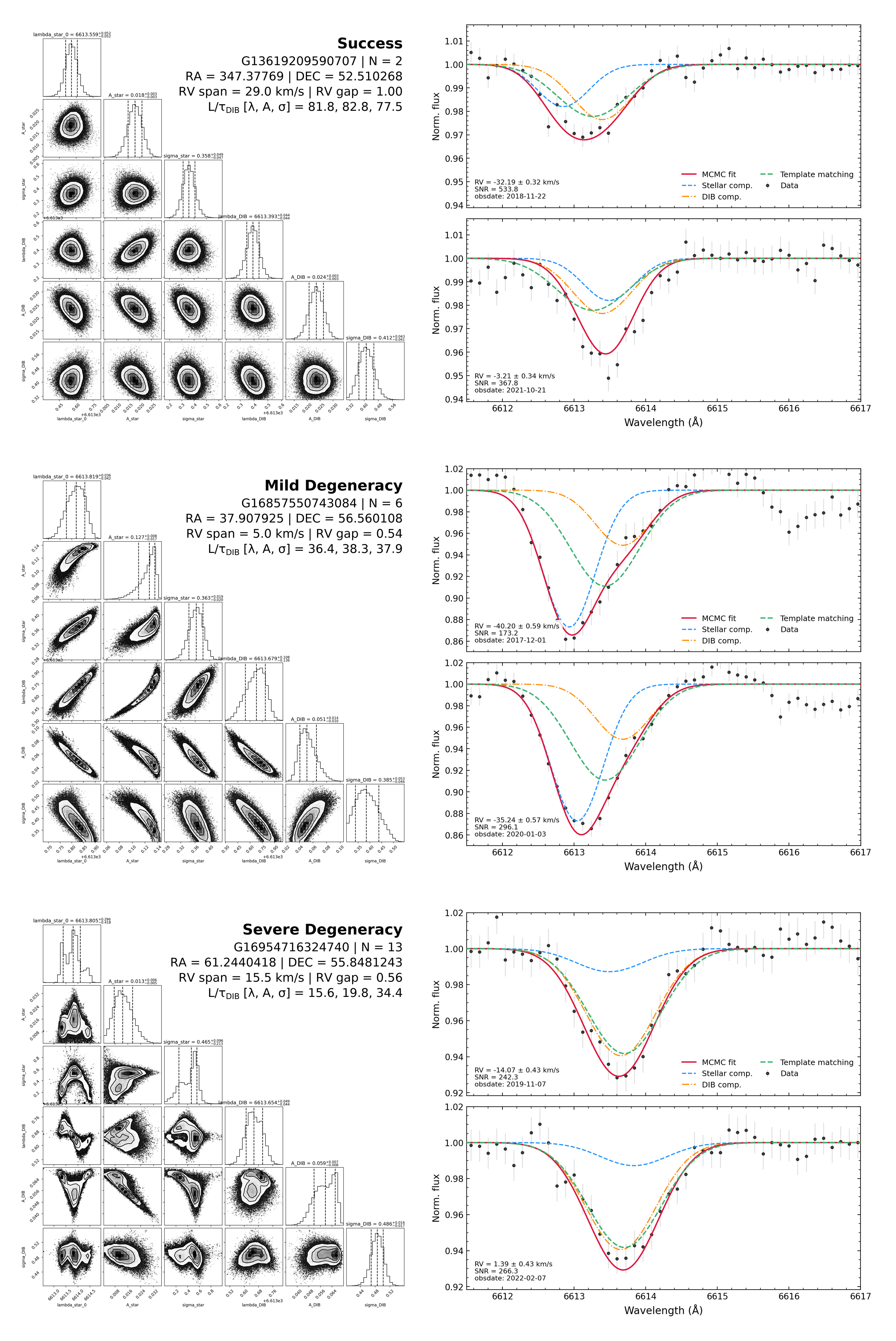}
    \caption{\label{fig:real_decomposition}\small MCMC-RVSD fits for the three degeneracy classes (Success / Mild Degeneracy / Severe Degeneracy), compared with the template-matching method. The left column shows the corner plots of the three classes, classified by the DIB autocorrelation time $L/\tau_{\mathrm{DIB}}$. The right column shows, for each target, the spectra of the two epochs with the most widely separated RVs: the scatter points with error bars represent the observed data, the red solid line is the best-fit model, the blue dashed and orange dash-dotted lines are the stellar and DIB components of the MCMC-RVSD fit, respectively, and the green dashed line is the DIB component inferred from template matching.}
\end{figure*}

We further manually inspected the posterior distributions in corner plots to verify the classification. As shown in the left of Figure~\ref{fig:real_decomposition}, the ``Success" class typically shows near-circular posterior distributions, indicating well-constrained parameters. The ``Severe" class exhibits highly diffuse posteriors and sometimes bimodal structures caused by other stellar lines interference. The ``Mild" class shows diverse behaviors. Some has highly degenerate, flattened elliptical distributions due to poor RV spans, while others have anomalous multi-peak structures.

\begin{figure*}[htb]
    \centering
    \includegraphics[width=0.8\textwidth]{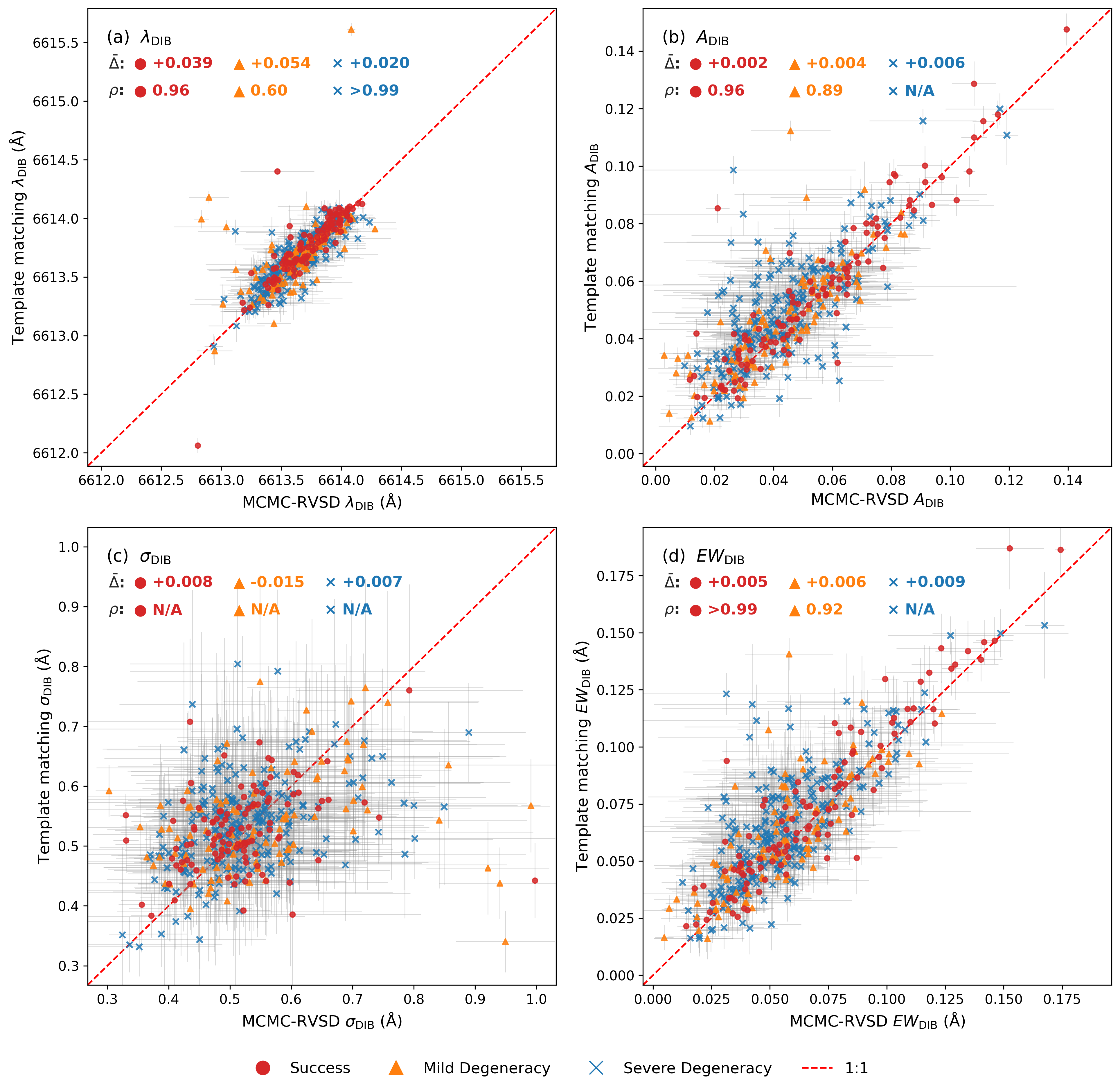}
    \caption{\label{fig:param_comparison}\small DIB $\lambda 6614$ parameters measured with MCMC-RVSD ($x$-axis) versus template matching ($y$-axis) for the three degeneracy classes. Error bars are the $1\sigma$ uncertainties of both methods and the dashed line is the 1:1 relation. In each panel the upper row gives the mean offset $\bar{\Delta} = \langle \mathrm{template}-\mathrm{MCMC}\rangle$ and the lower row the intrinsic correlation coefficient $\rho$ (errors-in-variables maximum-likelihood estimate, per-source errors on both axes). ``N/A'' marks cases where the quoted uncertainties account for the
    entire scatter, for which $\rho$ is not identifiable. Circles: Success;
    triangles: Mild Degeneracy; crosses: Severe Degeneracy.}
\end{figure*}

We therefore further compared the measurements of MCMC-RVSD with those obtained by template matching (Our sample is mainly composed of cool stars). Figure~\ref{fig:param_comparison} presents the measurements of the two methods as a scatter plot and quantifies their differences with two quantities. The first is the mean offset $\bar{\Delta} = \langle \mathrm{template}-\mathrm{MCMC}\rangle$, which indicates the direction in which the points depart from the one-to-one relation. The second is the intrinsic correlation coefficient $\rho$, which measures how consistent the two data sets are once their measurement errors are taken into account.

Since both axes carry measurement errors, $\rho$ cannot be obtained from the observed values directly and is instead estimated in an errors-in-variables framework. For each of the $N$ sources, whose measurements are $x_i$ (MCMC-RVSD) and $y_i$ (template matching), the $1\sigma$ uncertainties $e_{x,i}$ and $e_{y,i}$ on the two axes are known, so that the covariance matrix of the $i$-th measurement reads

\begin{equation}
\begin{aligned}
\ln\mathcal{L} &= -\frac{1}{2}\sum_{i=1}^{N} \Bigl[\ln\det V_i + \mathbf{d}_i^{\top} V_i^{-1}\mathbf{d}_i\Bigr],\\ V_i &= \begin{pmatrix} \sigma_X^{2}+e_{x,i}^{2} & \rho\,\sigma_X\sigma_Y\\ \rho\,\sigma_X\sigma_Y & \sigma_Y^{2}+e_{y,i}^{2} 
\end{pmatrix},
\end{aligned}
\label{eq:intrinsic_rho}
\end{equation}

The diagonal holds the measurement uncertainties added in quadrature to the intrinsic dispersions $\sigma_X$ and $\sigma_Y$ of the true values, while the off-diagonal term carries the very correlation we are after, $\rho \equiv \mathrm{cov}(X,Y)/(\sigma_X\sigma_Y)$. The log-likelihood $\ln\mathcal{L}$ of the measured pairs given the model is then maximised over the five free parameters $(\mu_X,\mu_Y,\sigma_X,\sigma_Y,\rho)$, where $\mathbf{d}_i = (x_i-\mu_X,\,y_i-\mu_Y)^{\top}$ is the deviation of source $i$ from the mean true values $\mu_X$ and $\mu_Y$. Because the noise is confined to the diagonal of $V_i$, the resulting $\rho$ is symmetric in the two methods and is not attenuated by them, unlike the ordinary Pearson coefficient computed from the observed values directly.

Figure~\ref{fig:param_comparison} shows that $\bar{\Delta} > 0$ for all parameters of the Success class, indicating that template matching generally measures larger DIB values than MCMC-RVSD, which may reflect an incomplete subtraction of the stellar lines in the template. For $\rho$, the correlation between the two methods behaves differently across the classes. For the Success class, the wavelength, the depth and the equivalent width agree closely with template matching, with $\rho \gtrsim 0.9$. As $\tau$ increases, the decomposition capability of MCMC-RVSD degrades, and the growing measurement errors make $\rho$ no longer identifiable for the Severe class. The central wavelength, however, is nearly unaffected by $\tau$. For $\sigma$, the precision provided by both methods is such that the quoted uncertainties already account for the entire spread of the measurements, so $\rho$ is not identifiable in any class. The anomalous $\rho$ found for the wavelength of the Mild class is caused by a few sources whose template-matching measurements are exceptionally large.

Overall, the $\tau$-based classification shows that the Success class reaches a higher measurement precision and nearly non-degenerate corner plots, and agrees most closely with the results of template matching. We also find that template matching generally measures larger DIB values than MCMC-RVSD, for the Success class larger by $0.039$\,\AA\ in the central wavelength and by $0.005$\,\AA\ ( $\approx8\%$) in the equivalent width.

\subsection{Correlation Analysis}
\label{subsec:correlation}

\begin{figure*}
   \centering
   \includegraphics[width=0.8\textwidth, angle=0]{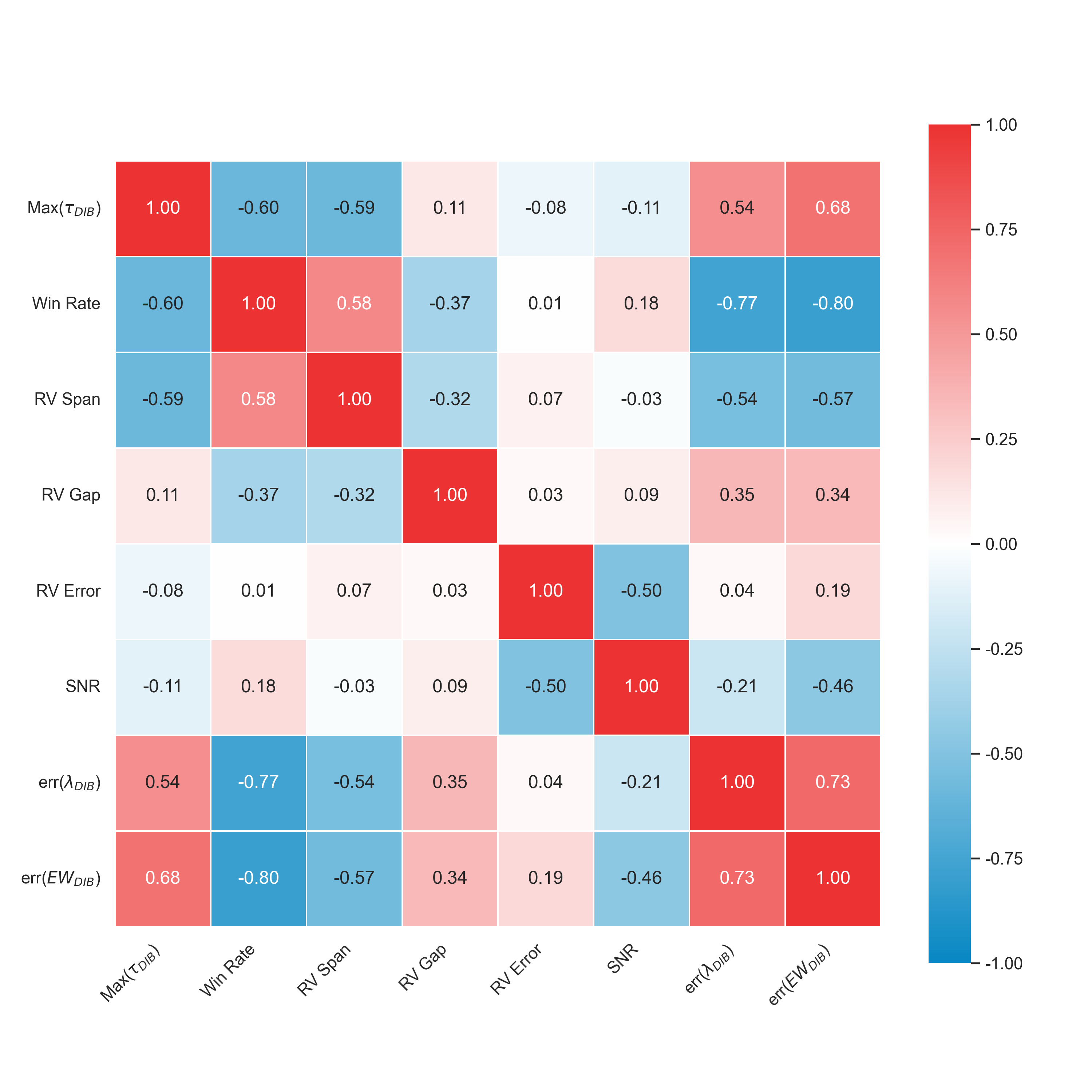}
   \caption{\small Spearman correlation matrix of observational metrics, stellar parameters, and fitting quality indicators. The matrix highlights several key findings: line fitting errors strongly anti-correlate with the algorithm's win rate and positively correlate with $\max(\tau_{\text{DIB}})$, validating our quality labels. Among observational parameters, RV span is the primary driver in reducing fitting errors by enhancing line separation, while RV gap provides supplementary sampling support. Furthermore, low-level RV errors show minimal disturbance, and SNR mainly limits parameter errors without dominating the MCMC convergence.}
   \label{fig:hot}
\end{figure*}

After proving the validity of the quality labels, we further analyzed the parameters affecting the fitting quality. We performed a Spearman correlation analysis on observational parameters (RV span, RV gap, RV error, and SNR), stellar parameters ($T_{\rm eff}$ and binary type), and the fitting results, which include the line fitting errors ($err(\lambda_{\rm DIB})$ and $err(EW_{\rm DIB})$),  $\max(\tau_{\mathrm{DIB}})$ and win rate. Here, RV span represents the range of the RV sampling, and RV gap denotes the RV sampling coverage. The ``win rate" refers to the fraction of cases where MCMC-RVSD outperforms template matching across the four DIB parameters for one source. 

Figure \ref{fig:hot} shows that the line fitting errors exhibit a moderate positive correlation with $\max(\tau_{\mathrm{DIB}})$ and a strong negative correlation with the win rate, further confirming the validity of using $\tau$ as the classification labels. 

As RV is the key parameter of the algorithm, its three parameters exhibited different impacts: an increase in RV span enhances the separation between stellar lines and interstellar medium lines, lowering $\tau$ and reducing line fitting errors; conversely, RV gap acts as supplementary sampling information that assists RV span in reducing errors, but it hardly affects the variation of $\tau$; RV error is almost uncorrelated with $\tau$ and fitting errors because the RV errors in our experimental sample are mostly below $5\,\text{km}\,\text{s}^{-1}$, which is consistent with the phenomenon observed in Section~\ref{subsec:Radial Velocity} where low-level RV errors cause minimal disturbance to the fitting errors. 

SNR shows a moderate negative correlation with the line fitting parameter errors but makes only a weak contribution to $\tau$ and the win rate. This may be because when SNR is greater than 50, it is no longer the primary factor affecting the convergence of the MCMC chain. The line fitting parameter errors exhibit a strong positive correlation with each other. Additionally, a weak correlation exists between RV gap and RV span, likely attributed to the fact that binaries with large radial velocity variations are more frequently sampled.

\subsection{Focus on RV Parameter}
To further investigate the impact of RV parameters, we analyzed the empirical cumulative distribution functions (ECDF) and kernel density estimations (KDE) of the RV span and RV gap for the three label classifications. 

We note a correlation coefficient of -0.32 between the RV gap and RV span, indicating an interplay between these parameters in the subsequent analysis. For instance, an RV gap of 1.0 implies only two sampling points; however, the uncertainty in sampling intervals can lead to varied classification labels. Although the combined effect of these two parameters complicates the analysis of $\tau$ classification and win rates, we still propose reasonable parameter thresholds for reference. 

\begin{figure*}
  \centering
  \includegraphics[width=0.99\textwidth]{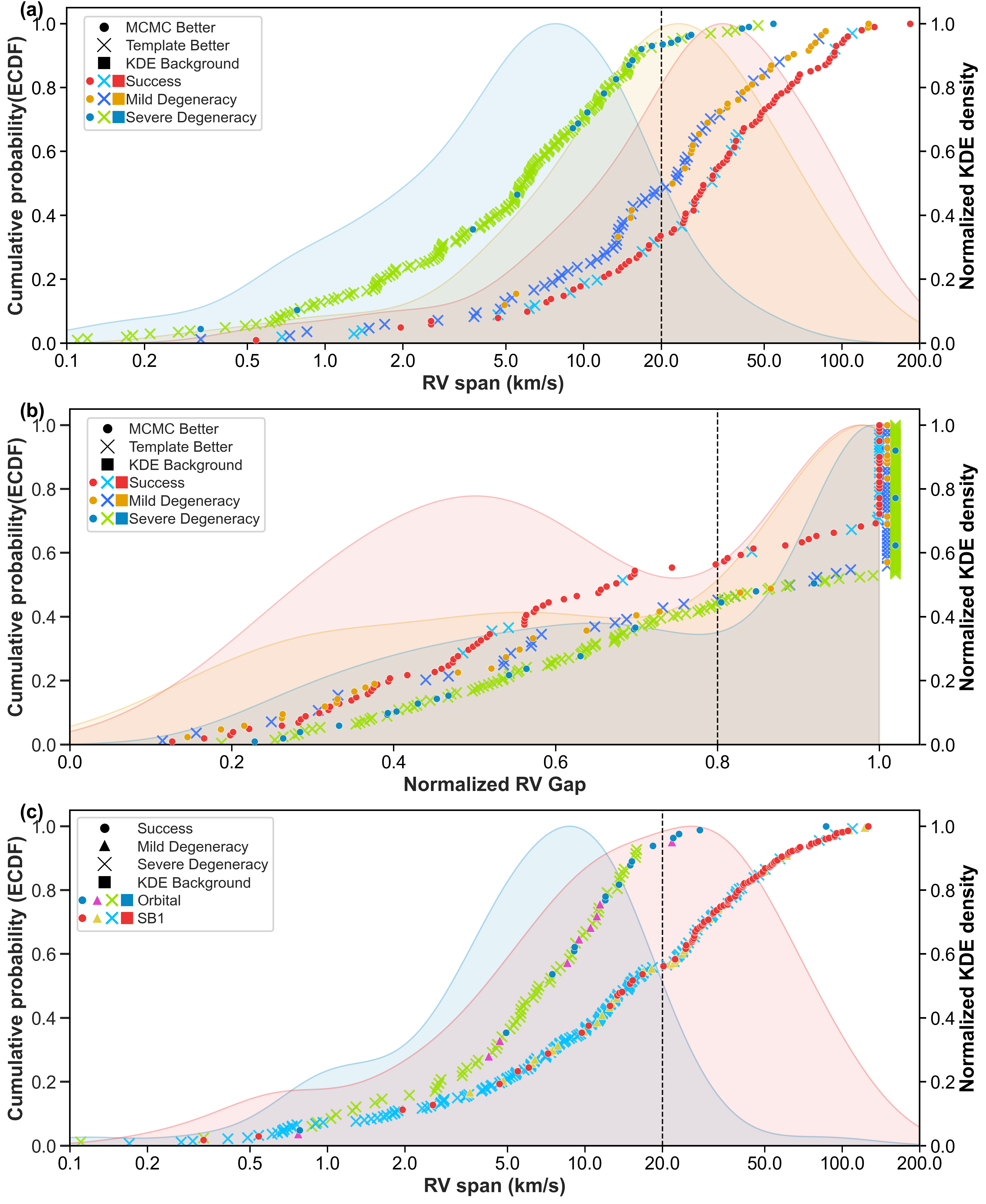}
  \caption{\label{fig:KDE}{\small {ECDF and KDE distributions of RV parameters and binary types. \textbf{(a)} RV span distributions across quality labels indicate that an RV span exceeding $20\,\text{km}\,\text{s}^{-1}$ significantly boosts the MCMC-RVSD algorithm's success rate. \textbf{(b)} RV gap distributions reveal that denser sampling coverage (e.g., RV gap $< 0.8$) enhances spectral decoupling capability, though robust results under sparse sampling (RV gap $= 1.0$) can occur if compensated by a large RV span. \textbf{(c)} Distributions for Orbital and SB1 binaries show that SB1 systems exhibit inherently larger RV spans due to detection biases, making their stellar lines naturally easier to separate from interstellar features.}}}
\end{figure*}

As shown in Figure \ref{fig:KDE} (a), the median density distributions of the RV span differ significantly across the three labels. The ``Success'' class exhibits the largest RV span, substantially exceeding that of the ``Severe'' class. Notably, when the RV span exceeds 20 $\rm km\,s^{-1}$, the measurement win rate of the MCMC-RVSD algorithm increases noticeably. 

Figure \ref{fig:KDE} (b) further reveals distinct density distributions between the ``Success'' and ``Severe'' classes. Using an RV gap of 0.8 as a reference threshold, $58\%$ of the ``Success'' samples demonstrate high sampling coverage, compared to only $45\%$ for the ``Severe'' class. This suggests that better sampling coverage enhances the algorithm's decoupling capability. For clarity, the scatter points for the three classes at RV gap = 1.0 have been slightly shifted. Observing the distribution of points (Success) and crosses (Severe), the latter dominate the RV gap = 1.0 region, accounting for $80\%$ of the population. The occasional appearance of points (Success) in this region was primarily driven by a large RV span, which boosts the algorithm's decoupling performance despite the sparse sampling.

Given that the template matching catalog focuses exclusively on cool stars and $T_{\rm eff}$ shows negligible correlation with other parameters, its influence is excluded from this analysis.  Additionally, the Gaia NSS catalog contains various binary types, with ``Orbital" and ``SB1" being the most numerous. Figure \ref{fig:KDE} (c) reveals that the median RV span of the Orbital binaries is much smaller than that of the SB1 systems. This was likely related to the observational detection bias for binary systems. The Orbital systems are often identified through long-term photometry or astrometry, whereas SB1 systems are explicitly confirmed via significant periodic radial velocity shifts in their spectra. Therefore, SB1 systems naturally exhibit larger RV spans, making their stellar lines much easier to separate from interstellar lines in the spectra.

\subsection{Summary and Catalog}

Finally, we compiled a complete catalog of DIB properties measured via the MCMC-RVSD algorithm. The dataset includes 434 unique sources with comprehensive fitting status labels, and parameter comparisons against traditional methods. The parameters detailed in the catalog consist of MCMC-derived profiles limits (central wavelength, depth, width), corresponding Gaia and LAMOST source identifiers, intrinsic stellar characteristics ($T_{\rm eff}$, $\log g$, [Fe/H]), and multiple-epoch RV distributions. A detailed description of the catalog columns is presented in Table \ref{tab:catalog_description}. Meanwhile, the catalog can be accessed at DataDOI: 10.57760/sciencedb.j00167.00054.

For DIB $\lambda 6614$ on LAMOST MRS, the RV span is the primary factor influencing the convergence of the algorithm, and an RV span greater than $20\,\text{km}\,\text{s}^{-1}$ is recommended. As supplementary information, the RV gap does not affect the convergence speed of the algorithm. An RV error below $5\,\text{km}\,\text{s}^{-1}$ and a SNR above 50 have almost no impact on the convergence of the MCMC chain. To improve the precision of the fitting parameters, high-SNR spectra are required to reduce noise interference, and a larger RV span is needed to accelerate the MCMC chain convergence. Finally, minimizing the RV gap enhances the decoupling accuracy of the algorithm; having more than three nearly uniformly distributed sampling points (RV gap $< 0.5$) can effectively reduce the fitting errors.

\begin{table*}
\centering
\caption{Catalog Description of the DIB $\lambda 6614$ Measurements with MCMC-RVSD on LAMOST MRS}
\label{tab:catalog_description}
\begin{tabular}{lcccl}
\hline
Index & Label & Format & Units & Description \\
\hline
1 & \verb|uid| & String & $\cdots$ & Unique source identifier \\
2 & \verb|gaia_source_id| & Integer & $\cdots$ & Gaia DR3 source\_id \\
3 & \verb|gaia_nss_solution_type| & String & $\cdots$ & Gaia Non-Single Star (NSS) solution type \\
4 & \verb|Label| & String & $\cdots$ & Label or classification of the target \\
5 & \verb|reduced_chi2| & Float & $\cdots$ & Reduced $\chi^2$ of the MCMC fitting \\
6 & \verb|mcmc_taus| & String & $\cdots$ & Autocorrelation times from MCMC sampling \\
7--12 & \verb|mcmc_[param]_[comp]| & Float & \AA$/\cdots$ & MCMC-RVSD fitted parameters for stellar and DIB \\
13--18 & \verb|mcmc_err_[param]_[comp]|& Float & \AA$/\cdots$ & MCMC-RVSD fitted parameters errors\\
19--22 & \verb|template_[param]_6614_mc|& Float & \AA$/\cdots$ & Template matching measurements\\
23--26 & \verb|template_[param]err_6614_mc|& Float & \AA$/\cdots$ & Template matching measurements errors \\
27--29 & \verb|[param]_lasp_list| & String & K/dex & Lists of LASP stellar parameters ($T_{\rm eff}$, $\log g$, [Fe/H]) \\
30--32 & \verb|lamost_[item]| & String & $\cdots$ & Lists of corresponding LAMOST observation info\\
33,35,39& \verb|[var]_distribution| & String & $\rm km\,s^{-1}/\cdots$& Ranges of RV, SNR, and RV errors \\
34,36,38& \verb|corresponding_[var]| & String & $\rm km\,s^{-1}/\cdots$& Lists of RVs, SNR values, and RV errors \\
37 & \verb|obsids| & String & $\cdots$ & Array of observation IDs \\
40--41 & \verb|rv_r0_err_[stat]| & Float & $\rm km\,s^{-1}$ & Statistical values (mean, median) of the RV errors \\
\hline
\end{tabular}
\end{table*}

\section{Conclusion and Discussion}
\label{sect:conclusion}
In this paper, we presented a new Bayesian algorithm (MCMC-RVSD) to accurately measure Diffuse Interstellar Bands (DIBs) in binary star spectra. By using radial velocity (RV) as the core parameter, our method successfully separates moving stellar lines from stationary DIBs.

We first verified the algorithm using mock spectra based on LAMOST medium-resolution data. Under ideal conditions ($R=7500$, $\text{SNR}=200$), the algorithm decoupled the DIB $\lambda$6614 and the nearby Fe I line with high precision. Wavelength precision reached $0.01\,\mathrm{\AA}$, while depth and width precisions reached $0.01$ and $0.01\,\mathrm{\AA}$. We also found that this method relies on current prior empirical boundaries, making it optimally designed for spectra with a resolution of $R=3000$ to $7500$.

We then applied the MCMC-RVSD method to 434 real binary sources from the Gaia DR3 and LAMOST surveys. We introduced the MCMC autocorrelation time ($\tau_{\mathrm{DIB}}$) as a reliable metric to evaluate the fit quality. This effectively classified the results into ``Success'', ``Mild Degeneracy'', and ``Severe Degeneracy''.

For DIB $\lambda 6614$ at typical profile ($A_{\text{DIB}}=0.049$,\  $\sigma_{\text{DIB}}=0.546\,\rm{\AA}$,\ $E(B-V)\approx0.3$ mag)  on LAMOST MRS, validation tests and correlation analyses revealed the key factors for a successful fit:
\begin{itemize}
    \item \textbf{RV Span}: A span $>20\,\mathrm{km\,s^{-1}}$ is highly recommended because it physically separates the lines. Single-lined spectroscopic binaries (SB1) are ideal targets as they naturally provide large RV spans.
    \item \textbf{RV Gap}: Denser sampling (smaller RV gaps) helps, but a large RV span alone can ensure robust measurements even with sparse sampling.
    \item \textbf{Noise and Errors}: The algorithm is highly robust. When $\text{SNR} > 50$ and RV errors are $< 5\,\text{km}\,\text{s}^{-1}$, they hardly affect the MCMC convergence.
\end{itemize}

Based on these rules, we published a high-precision catalog of DIB measurements for the 434 binary systems. Under the ideal condition, our method ultimately provides much smaller parameter errors than traditional template matching.

For the complex spectral situations encountered in real data, we tested contamination from multiple stellar lines, non-Gaussian DIB profiles, and other DIBs
in the mock data. Once the corresponding model is adopted, the input values of the mock spectra fall within the $3\sigma$ range of the posterior distributions. For even more complex DIBs and interstellar-medium lines, for which no prior information is available, the quality labels of Section~\ref{subsec:label} provide the final safeguard.

In the future, we will develop an open-source Python package at \url{https://github.com/hss3255/mcmc-rvsd}, enabling the community measure and analyze other interstellar lines in binary systems.

\begin{acknowledgements}
The authors acknowledge the support by the NationalKey Research and Development Program of China (2025YFF0511001).  This work is also supported by the National Astronomical Observatories of the ChineseAcademy of Sciences, under Project No. E4TQ680101. Thanks the support of the China Manned Space Program (Grant No. CMS-CSST-2025-A19).

The catalog of this work will be supplied to Sience Data Bnak with Data DOI (10.57760/sciencedb.j00167.00054). This work made use of the data from LAMOST (Large Sky Area Multi-Object Fiber Spectroscopic Telescope, also known as the Guoshoujing Telescope) (https://cstr.cn/31118.02.LAMOST). LAMOST is a Chinese national mega-science facility, operated by National Astronomical Observatories, Chinese Academy of Sciences.
\end{acknowledgements}

\appendix                  

\bibliographystyle{raa}
\bibliography{ms2026-0385}

@ARTICLE{1995ARA&A..33...19H,
       author = {{Herbig}, G.~H.},
        title = "{The Diffuse Interstellar Bands}",
      journal = {\araa},
         year = 1995,
        month = jan,
       volume = {33},
        pages = {19-74},
          doi = {10.1146/annurev.aa.33.090195.000315},
       adsurl = {https://ui.adsabs.harvard.edu/abs/1995ARA&A..33...19H}
}

@ARTICLE{1922LicOB..10..146H,
       author = {{Heger}, Mary Lea},
        title = "{The spectra of certain class B stars in the regions 5630A-6680A and 3280A-3380A}",
      journal = {Lick Observatory Bulletin},
         year = 1922,
        month = jan,
       volume = {10},
       number = {337},
        pages = {146-147},
       adsurl = {https://ui.adsabs.harvard.edu/abs/1922LicOB..10..146H}
}

@ARTICLE{2022ApJS..262....2H,
       author = {{Hamano}, Satoshi and {Kobayashi}, Naoto and {Kawakita}, Hideyo and {Takenaka}, Keiichi and {Ikeda}, Yuji and {Matsunaga}, Noriyuki and {Kondo}, Sohei and {Sameshima}, Hiroaki and {Fukue}, Kei and {Otsubo}, Shogo and {Arai}, Akira and {Yasui}, Chikako and {Kobayashi}, Hitomi and {Bono}, Giuseppe and {Saviane}, Ivo},
        title = "{Survey of Near-infrared Diffuse Interstellar Bands in Y and J Bands. I. Newly Identified Bands}",
      journal = {\apjs},
         year = 2022,
        month = sep,
       volume = {262},
       number = {1},
          eid = {2},
        pages = {2},
          doi = {10.3847/1538-4365/ac7567},
archivePrefix = {arXiv},
       eprint = {2206.03131},
 primaryClass = {astro-ph.GA},
       adsurl = {https://ui.adsabs.harvard.edu/abs/2022ApJS..262....2H}
}

@ARTICLE{2019ApJ...878..151F,
       author = {{Fan}, Haoyu and {Hobbs}, L.~M. and {Dahlstrom}, Julie A. and {Welty}, Daniel E. and {York}, Donald G. and {Rachford}, Brian and {Snow}, Theodore P. and {Sonnentrucker}, Paule and {Baskes}, Noah and {Zhao}, Gang},
        title = "{The Apache Point Observatory Catalog of Optical Diffuse Interstellar Bands}",
      journal = {\apj},
         year = 2019,
        month = jun,
       volume = {878},
       number = {2},
          eid = {151},
        pages = {151},
          doi = {10.3847/1538-4357/ab1b74},
archivePrefix = {arXiv},
       eprint = {1905.05962},
 primaryClass = {astro-ph.GA},
       adsurl = {https://ui.adsabs.harvard.edu/abs/2019ApJ...878..151F}
}

@ARTICLE{1975ApJ...196..129H,
       author = {{Herbig}, G.~H.},
        title = "{The diffuse interstellar bands. The region 4400 - 6850 {\r{A}}.}",
      journal = {\apj},
         year = 1975,
        month = feb,
       volume = {196},
        pages = {129-160},
          doi = {10.1086/153400},
       adsurl = {https://ui.adsabs.harvard.edu/abs/1975ApJ...196..129H}
}

@ARTICLE{2004ApJ...611L.113C,
       author = {{Cami}, J. and {Salama}, F. and {Jim{\'e}nez-Vicente}, J. and {Galazutdinov}, G.~A. and {Kre{\l}owski}, J.},
        title = "{The Rotational Excitation Temperature of the {\ensuremath{\lambda}}6614 Diffuse Interstellar Band Carrier}",
      journal = {\apjl},
         year = 2004,
        month = aug,
       volume = {611},
       number = {2},
        pages = {L113-L116},
          doi = {10.1086/423991},
archivePrefix = {arXiv},
       eprint = {astro-ph/0407415},
 primaryClass = {astro-ph},
       adsurl = {https://ui.adsabs.harvard.edu/abs/2004ApJ...611L.113C}
}

@ARTICLE{2019ApJ...875L..28C,
       author = {{Cordiner}, M.~A. and {Linnartz}, H. and {Cox}, N.~L.~J. and {Cami}, J. and {Najarro}, F. and {Proffitt}, C.~R. and {Lallement}, R. and {Ehrenfreund}, P. and {Foing}, B.~H. and {Gull}, T.~R. and {Sarre}, P.~J. and {Charnley}, S.~B.},
        title = "{Confirming Interstellar C$_{60}$ $^{+}$ Using the Hubble Space Telescope}",
      journal = {\apjl},
         year = 2019,
        month = apr,
       volume = {875},
       number = {2},
          eid = {L28},
        pages = {L28},
          doi = {10.3847/2041-8213/ab14e5},
archivePrefix = {arXiv},
       eprint = {1904.08821},
 primaryClass = {astro-ph.GA},
       adsurl = {https://ui.adsabs.harvard.edu/abs/2019ApJ...875L..28C}
}

@ARTICLE{2025MNRAS.543.1429C,
       author = {{Chang}, Chih-Yuan and {Lan}, Ting-Wen},
        title = "{A deep search for diffuse interstellar band 4430 {\r{A}} in the circumgalactic medium}",
      journal = {\mnras},
         year = 2025,
        month = oct,
       volume = {543},
       number = {2},
        pages = {1429-1434},
          doi = {10.1093/mnras/staf1560},
archivePrefix = {arXiv},
       eprint = {2501.07082},
 primaryClass = {astro-ph.GA},
       adsurl = {https://ui.adsabs.harvard.edu/abs/2025MNRAS.543.1429C}
}

@ARTICLE{2025RAA....25i7001M,
       author = {{Ma}, Xiao-Xiao and {Luo}, A.-Li and {Chen}, Jian-Jun and {Chen}, Jing and {Liang}, Jun-Chao},
        title = "{Measurements of the Diffuse Interstellar Bands at 5780, 5797, and 6614 {\r{A}} in the Hot Stellar Spectra of the LAMOST LRS DR10}",
      journal = {Research in Astronomy and Astrophysics},
         year = 2025,
        month = sep,
       volume = {25},
       number = {9},
          eid = {097001},
        pages = {097001},
          doi = {10.1088/1674-4527/ade58d},
archivePrefix = {arXiv},
       eprint = {2506.14346},
 primaryClass = {astro-ph.GA},
       adsurl = {https://ui.adsabs.harvard.edu/abs/2025RAA....25i7001M}
}

@ARTICLE{2014MNRAS.439.3139Z,
       author = {{Zhu}, Guangtun and {M{\'e}nard}, Brice and {Bizyaev}, Dmitry and {Brewington}, Howard and {Ebelke}, Garrett and {Ho}, Shirley and {Kinemuchi}, Karen and {Malanushenko}, Viktor and {Malanushenko}, Elena and {Marchante}, Moses and {More}, Surhud and {Oravetz}, Daniel and {Pan}, Kaike and {Petitjean}, Patrick and {Simmons}, Audrey},
        title = "{The large-scale distribution of cool gas around luminous red galaxies}",
      journal = {\mnras},
         year = 2014,
        month = apr,
       volume = {439},
       number = {3},
        pages = {3139-3155},
          doi = {10.1093/mnras/stu186},
archivePrefix = {arXiv},
       eprint = {1309.7660},
 primaryClass = {astro-ph.CO},
       adsurl = {https://ui.adsabs.harvard.edu/abs/2014MNRAS.439.3139Z}
}

@ARTICLE{2018ApJ...866...36L,
       author = {{Lan}, Ting-Wen and {Mo}, Houjun},
        title = "{The Circumgalactic Medium of eBOSS Emission Line Galaxies: Signatures of Galactic Outflows in Gas Distribution and Kinematics}",
      journal = {\apj},
         year = 2018,
        month = oct,
       volume = {866},
       number = {1},
          eid = {36},
        pages = {36},
          doi = {10.3847/1538-4357/aadc08},
archivePrefix = {arXiv},
       eprint = {1806.05786},
 primaryClass = {astro-ph.GA},
       adsurl = {https://ui.adsabs.harvard.edu/abs/2018ApJ...866...36L}
}

@ARTICLE{2002ApJS..139..439R,
       author = {{Redfield}, Seth and {Linsky}, Jeffrey L.},
        title = "{The Structure of the Local Interstellar Medium. I. High-Resolution Observations of Fe II, Mg II, and Ca II toward Stars within 100 Parsecs}",
      journal = {\apjs},
         year = 2002,
        month = apr,
       volume = {139},
       number = {2},
        pages = {439-465},
          doi = {10.1086/338650},
       adsurl = {https://ui.adsabs.harvard.edu/abs/2002ApJS..139..439R}
}

@ARTICLE{1993A&A...278..549B,
       author = {{Bertin}, P. and {Lallement}, R. and {Ferlet}, R. and {Vidal-Madjar}, A.},
        title = "{The NaI/Ca II ratio in the local interstellar medium.}",
      journal = {\aap},
         year = 1993,
        month = nov,
       volume = {278},
        pages = {549-560},
       adsurl = {https://ui.adsabs.harvard.edu/abs/1993A&A...278..549B}
}

@ARTICLE{1996A&A...307L..25E,
       author = {{Ehrenfreund}, P. and {Foing}, B.~H.},
        title = "{Resolved profiles of diffuse interstellar bands: evidence fou rotational contours of gas phase molecules.}",
      journal = {\aap},
         year = 1996,
        month = mar,
       volume = {307},
        pages = {L25-L28},
       adsurl = {https://ui.adsabs.harvard.edu/abs/1996A&A...307L..25E}
}

@ARTICLE{2011ApJ...727...33F,
       author = {{Friedman}, Scott D. and {York}, Donald G. and {McCall}, Benjamin J. and {Dahlstrom}, Julie and {Sonnentrucker}, Paule and {Welty}, Daniel E. and {Drosback}, Meredith M. and {Hobbs}, L.~M. and {Rachford}, Brian L. and {Snow}, Theodore P.},
        title = "{Studies of Diffuse Interstellar Bands V. Pairwise Correlations of Eight Strong DIBs and Neutral Hydrogen, Molecular Hydrogen, and Color Excess}",
      journal = {\apj},
         year = 2011,
        month = jan,
       volume = {727},
       number = {1},
          eid = {33},
        pages = {33},
          doi = {10.1088/0004-637X/727/1/33},
archivePrefix = {arXiv},
       eprint = {1011.2951},
 primaryClass = {astro-ph.GA},
       adsurl = {https://ui.adsabs.harvard.edu/abs/2011ApJ...727...33F}
}

@ARTICLE{2022A&A...662A..81E,
       author = {{Ebenbichler}, A. and {Postel}, A. and {Przybilla}, N. and {Seifahrt}, A. and {We{\ss}mayer}, D. and {Kausch}, W. and {Firnstein}, M. and {Butler}, K. and {Kaufer}, A. and {Linnartz}, H.},
        title = "{CRIRES high-resolution near-infrared spectroscopy of diffuse interstellar band profiles. Detection of 12 new DIBs in the YJ band and the introduction of a combined ISM sight line and stellar analysis approach}",
      journal = {\aap},
         year = 2022,
        month = jun,
       volume = {662},
          eid = {A81},
        pages = {A81},
          doi = {10.1051/0004-6361/202142990},
archivePrefix = {arXiv},
       eprint = {2203.12562},
 primaryClass = {astro-ph.GA},
       adsurl = {https://ui.adsabs.harvard.edu/abs/2022A&A...662A..81E}
}

@ARTICLE{2010ApJ...708.1628M,
       author = {{McCall}, Benjamin J. and {Drosback}, Meredith M. and {Thorburn}, Julie A. and {York}, Donald G. and {Friedman}, Scott D. and {Hobbs}, L.~M. and {Rachford}, Brian L. and {Snow}, Theodore P. and {Sonnentrucker}, Paule and {Welty}, Daniel E.},
        title = "{Studies of the Diffuse Interstellar Bands. IV. The Nearly Perfect Correlation Between {\ensuremath{\lambda}}{\ensuremath{\lambda}}6196.0 and 6613.6}",
      journal = {\apj},
         year = 2010,
        month = jan,
       volume = {708},
       number = {2},
        pages = {1628-1638},
          doi = {10.1088/0004-637X/708/2/1628},
       adsurl = {https://ui.adsabs.harvard.edu/abs/2010ApJ...708.1628M}
}

@ARTICLE{2025A&A...702A..58S,
       author = {{Shi}, Ji-Wei and {Chen}, Jian-Jun and {Zhao}, He and {Ma}, Xiao-Xiao and {Luo}, A. -Li},
        title = "{Diffuse interstellar band {\ensuremath{\lambda}}6614 as a distance tracer for the Perseus, Taurus, and Orion molecular clouds: Feasibility and limitations}",
      journal = {\aap},
         year = 2025,
        month = oct,
       volume = {702},
          eid = {A58},
        pages = {A58},
          doi = {10.1051/0004-6361/202553656},
       adsurl = {https://ui.adsabs.harvard.edu/abs/2025A&A...702A..58S}
}

@ARTICLE{2012RAA....12.1197C,
       author = {{Cui}, Xiang-Qun and {Zhao}, Yong-Heng and {Chu}, Yao-Quan and {Li}, Guo-Ping and {Li}, Qi and {Zhang}, Li-Ping and {Su}, Hong-Jun and {Yao}, Zheng-Qiu and {Wang}, Ya-Nan and {Xing}, Xiao-Zheng and {Li}, Xin-Nan and {Zhu}, Yong-Tian and {Wang}, Gang and {Gu}, Bo-Zhong and {Luo}, A.-Li and {Xu}, Xin-Qi and {Zhang}, Zhen-Chao and {Liu}, Gen-Rong and {Zhang}, Hao-Tong and {Yang}, De-Hua and {Cao}, Shu-Yun and {Chen}, Hai-Yuan and {Chen}, Jian-Jun and {Chen}, Kun-Xin and {Chen}, Ying and {Chu}, Jia-Ru and {Feng}, Lei and {Gong}, Xue-Fei and {Hou}, Yong-Hui and {Hu}, Hong-Zhuan and {Hu}, Ning-Sheng and {Hu}, Zhong-Wen and {Jia}, Lei and {Jiang}, Fang-Hua and {Jiang}, Xiang and {Jiang}, Zi-Bo and {Jin}, Ge and {Li}, Ai-Hua and {Li}, Yan and {Li}, Ye-Ping and {Liu}, Guan-Qun and {Liu}, Zhi-Gang and {Lu}, Wen-Zhi and {Mao}, Yin-Dun and {Men}, Li and {Qi}, Yong-Jun and {Qi}, Zhao-Xiang and {Shi}, Huo-Ming and {Tang}, Zheng-Hong and {Tao}, Qing-Sheng and {Wang}, Da-Qi and {Wang}, Dan and {Wang}, Guo-Min and {Wang}, Hai and {Wang}, Jia-Ning and {Wang}, Jian and {Wang}, Jian-Ling and {Wang}, Jian-Ping and {Wang}, Lei and {Wang}, Shu-Qing and {Wang}, You and {Wang}, Yue-Fei and {Xu}, Ling-Zhe and {Xu}, Yan and {Yang}, Shi-Hai and {Yu}, Yong and {Yuan}, Hui and {Yuan}, Xiang-Yan and {Zhai}, Chao and {Zhang}, Jing and {Zhang}, Yan-Xia and {Zhang}, Yong and {Zhao}, Ming and {Zhou}, Fang and {Zhou}, Guo-Hua and {Zhu}, Jie and {Zou}, Si-Cheng},
        title = "{The Large Sky Area Multi-Object Fiber Spectroscopic Telescope (LAMOST)}",
      journal = {Research in Astronomy and Astrophysics},
         year = 2012,
        month = sep,
       volume = {12},
       number = {9},
        pages = {1197-1242},
          doi = {10.1088/1674-4527/12/9/003},
       adsurl = {https://ui.adsabs.harvard.edu/abs/2012RAA....12.1197C}
}

@ARTICLE{2012RAA....12.1243L,
       author = {{Luo}, A. -Li and {Zhang}, Hao-Tong and {Zhao}, Yong-Heng and {Zhao}, Gang and {Cui}, Xiang-Qun and {Li}, Guo-Ping and {Chu}, Yao-Quan and {Shi}, Jian-Rong and {Wang}, Gang and {Zhang}, Jian-Nan and {Bai}, Zhong-Rui and {Chen}, Xiao-Yan and {Wang}, Feng-Fei and {Guo}, Yan-Xin and {Chen}, Jian-Jun and {Du}, Bing and {Kong}, Xiao and {Lei}, Ya-Juan and {Li}, Yin-Bi and {Song}, Yi-Han and {Wu}, Yue and {Zhang}, Yan-Xia and {Zhou}, Xin-Lin and {Zuo}, Fang and {Du}, Peng and {He}, Lin and {Hou}, Wen and {Dong}, Yi-Qiao and {Li}, Jian and {Li}, Guang-Wei and {Li}, Shuang and {Song}, Jing and {Tian}, Yuan and {Wang}, Meng-Xin and {Wu}, Ke-Fei and {Yang}, Hui-Qin and {Yuan}, Hai-Long and {Cao}, Shu-Yun and {Chen}, Hai-Yuan and {Chen}, Kun-Xin and {Chen}, Ying and {Chu}, Jia-Ru and {Feng}, Lei and {Gong}, Xue-Fei and {Gu}, Bo-Zhong and {Hou}, Yong-Hui and {Huo}, Zhi-Ying and {Hu}, Hong-Zhuan and {Hu}, Ning-Sheng and {Hu}, Zhong-Wen and {Jia}, Lei and {Jiang}, Fang-Hua and {Jiang}, Xiang and {Jiang}, Zi-Bo and {Jin}, Ge and {Li}, Ai-Hua and {Li}, Qi and {Li}, Xin-Nan and {Li}, Yan and {Li}, Ye-Ping and {Liu}, Gen-Rong and {Liu}, Guan-Qun and {Liu}, Zhi-Gang and {Lu}, Qi-Shuai and {Lu}, Wen-Zhi and {Luo}, Yu and {Mao}, Yin-Dun and {Men}, Li and {Ni}, Ji-Jun and {Qi}, Yong-Jun and {Qi}, Zhao-Xiang and {Shi}, Huo-Ming and {Su}, Ding-Qiang and {Sun}, Shi-Wei and {Su}, Hong-Jun and {Tang}, Zheng-Hong and {Tao}, Qing-Sheng and {Tu}, Liang-Ping and {Wang}, Da-Qing and {Wang}, Dan and {Wang}, Guo-Min and {Wang}, Hai and {Wang}, Jia-Ning and {Wang}, Jian and {Wang}, Jian-Ling and {Wang}, Jian-Ping and {Wang}, Lei and {Wang}, Shou-Guan and {Wang}, Shu-Qing and {Wang}, Ya-Nan and {Wang}, You and {Wang}, Yue-Fei and {Wei}, Ming-Zhi and {Xue}, Xiang-Xiang and {Xing}, Xiao-Zheng and {Xu}, Ling-Zhe and {Xu}, Xin-Qi and {Xu}, Yan and {Yang}, De-Hua and {Yang}, Shi-Hai and {Yao}, Zheng-Qiu and {Yu}, Yong and {Yuan}, Hui and {Zhai}, Chao and {Zhang}, En-Peng and {Zhang}, Jing and {Zhang}, Li-Ping and {Zhang}, Wei and {Zhang}, Yong and {Zhang}, Zhen-Chao and {Zhao}, Ming and {Zhou}, Fang and {Zhu}, Yong-Tian and {Zhu}, Jie and {Zou}, Si-Cheng},
        title = "{Data release of the LAMOST pilot survey}",
      journal = {Research in Astronomy and Astrophysics},
         year = 2012,
        month = sep,
       volume = {12},
       number = {9},
        pages = {1243-1246},
          doi = {10.1088/1674-4527/12/9/004},
       adsurl = {https://ui.adsabs.harvard.edu/abs/2012RAA....12.1243L}
}

@ARTICLE{2012RAA....12..723Z,
       author = {{Zhao}, Gang and {Zhao}, Yong-Heng and {Chu}, Yao-Quan and {Jing}, Yi-Peng and {Deng}, Li-Cai},
        title = "{LAMOST spectral survey {\textemdash} An overview}",
      journal = {Research in Astronomy and Astrophysics},
         year = 2012,
        month = jul,
       volume = {12},
       number = {7},
        pages = {723-734},
          doi = {10.1088/1674-4527/12/7/002},
       adsurl = {https://ui.adsabs.harvard.edu/abs/2012RAA....12..723Z}
}

@ARTICLE{2019ApJS..244...27W,
       author = {{Wang}, R. and {Luo}, A.-L. and {Chen}, J.-J. and {Bai}, Z.-R. and {Chen}, L. and {Chen}, X.-F. and {Dong}, S.-B. and {Du}, B. and {Fu}, J.-N. and {Han}, Z.-W. and {Hou}, J.-L. and {Hou}, Y.-H. and {Hou}, W. and {Jiang}, D.-K. and {Kong}, X. and {Li}, L.-F. and {Liu}, C. and {Liu}, J.-M. and {Qin}, L. and {Shi}, J.-R. and {Tian}, H. and {Wu}, H. and {Wu}, C.-J. and {Xie}, J.-W. and {Zhang}, H.-T. and {Zhang}, S. and {Zhao}, G. and {Zhao}, Y.-H. and {Zhong}, J. and {Zong}, W.-K. and {Zuo}, F.},
        title = "{Properties of Radial Velocities Measurement Based on LAMOST-II Medium-resolution Spectroscopic Observations}",
      journal = {\apjs},
         year = 2019,
        month = oct,
       volume = {244},
       number = {2},
          eid = {27},
        pages = {27},
          doi = {10.3847/1538-4365/ab3cc0},
archivePrefix = {arXiv},
       eprint = {1908.04773},
 primaryClass = {astro-ph.SR},
       adsurl = {https://ui.adsabs.harvard.edu/abs/2019ApJS..244...27W}
}

@Misc{NIST_ASD,
author = {A.~Kramida and {Yu.~Ralchenko} and
J.~Reader and {and NIST ASD Team}},
HOWPUBLISHED = {{NIST Atomic Spectra Database
(ver. 5.12), [Online]. Available:
{\tt{https://physics.nist.gov/asd}} [2026, January 15].
National Institute of Standards and Technology,
Gaithersburg, MD.}},
year = {2024},
}

@BOOK{1966sst..book.....M,
       author = {{Moore}, Charlotte E. and {Minnaert}, M.~G.~J. and {Houtgast}, J.},
        title = "{The solar spectrum 2935 A to 8770 A}",
         year = 1966,
       adsurl = {https://ui.adsabs.harvard.edu/abs/1966sst..book.....M}
}

@ARTICLE{2011CaJPh..89..395L,
       author = {{Lobel}, Alex},
        title = "{Oscillator strength measurements of atomic absorption lines from stellar spectra}",
      journal = {Canadian Journal of Physics},
         year = 2011,
        month = may,
       volume = {89},
        pages = {395-402},
          doi = {10.1139/p11-030},
archivePrefix = {arXiv},
       eprint = {1012.5243},
 primaryClass = {astro-ph.SR},
       adsurl = {https://ui.adsabs.harvard.edu/abs/2011CaJPh..89..395L}
}

@book{2007nras.book.....P,
    author = {{Press}, William H. and {Teukolsky}, Saul A. and {Vetterling}, William T. and {Flannery}, Brian P.},
  title     = {Numerical Recipes: The Art of Scientific Computing},
  edition   = {3},
  year      = {2007},
  publisher = {Cambridge University Press},
  address   = {Cambridge, UK},
  isbn      = {978-0-521-88068-8},
  url       = {https://ui.adsabs.harvard.edu/abs/2007nras.book.....P},
  note      = {Provided by the SAO/NASA Astrophysics Data System},
}

@ARTICLE{2013PASP..125..306F,
       author = {{Foreman-Mackey}, Daniel and {Hogg}, David W. and {Lang}, Dustin and {Goodman}, Jonathan},
        title = "{emcee: The MCMC Hammer}",
      journal = {\pasp},
         year = 2013,
        month = mar,
       volume = {125},
       number = {925},
        pages = {306},
          doi = {10.1086/670067},
archivePrefix = {arXiv},
       eprint = {1202.3665},
 primaryClass = {astro-ph.IM},
       adsurl = {https://ui.adsabs.harvard.edu/abs/2013PASP..125..306F}
}

@ARTICLE{2024ExA....57...12K,
       author = {{Koen}, Chris},
        title = "{How many random observations are needed for good phase coverage of a periodic source?}",
      journal = {Experimental Astronomy},
         year = 2024,
        month = mar,
       volume = {57},
       number = {2},
          eid = {12},
        pages = {12},
          doi = {10.1007/s10686-024-09931-1},
       adsurl = {https://ui.adsabs.harvard.edu/abs/2024ExA....57...12K}
}

@ARTICLE{2023A&A...674A...1G,
       author = {{Gaia Collaboration} and {Vallenari}, A. and {Brown}, A.~G.~A. and {Prusti}, T. and {de Bruijne}, J.~H.~J. and {Arenou}, F. and {Babusiaux}, C. and {Biermann}, M. and {Creevey}, O.~L. and {Ducourant}, C. and {Evans}, D.~W. and {Eyer}, L. and {Guerra}, R. and {Hutton}, A. and {Jordi}, C. and {Klioner}, S.~A. and {Lammers}, U.~L. and {Lindegren}, L. and {Luri}, X. and {Mignard}, F. and {Panem}, C. and {Pourbaix}, D. and {Randich}, S. and {Sartoretti}, P. and {Soubiran}, C. and {Tanga}, P. and {Walton}, N.~A. and {Bailer-Jones}, C.~A.~L. and {Bastian}, U. and {Drimmel}, R. and {Jansen}, F. and {Katz}, D. and {Lattanzi}, M.~G. and {van Leeuwen}, F. and {Bakker}, J. and {Cacciari}, C. and {Casta{\~n}eda}, J. and {De Angeli}, F. and {Fabricius}, C. and {Fouesneau}, M. and {Fr{\'e}mat}, Y. and {Galluccio}, L. and {Guerrier}, A. and {Heiter}, U. and {Masana}, E. and {Messineo}, R. and {Mowlavi}, N. and {Nicolas}, C. and {Nienartowicz}, K. and {Pailler}, F. and {Panuzzo}, P. and {Riclet}, F. and {Roux}, W. and {Seabroke}, G.~M. and {Sordo}, R. and {Th{\'e}venin}, F. and {Gracia-Abril}, G. and {Portell}, J. and {Teyssier}, D. and {Altmann}, M. and {Andrae}, R. and {Audard}, M. and {Bellas-Velidis}, I. and {Benson}, K. and {Berthier}, J. and {Blomme}, R. and {Burgess}, P.~W. and {Busonero}, D. and {Busso}, G. and {C{\'a}novas}, H. and {Carry}, B. and {Cellino}, A. and {Cheek}, N. and {Clementini}, G. and {Damerdji}, Y. and {Davidson}, M. and {de Teodoro}, P. and {Nu{\~n}ez Campos}, M. and {Delchambre}, L. and {Dell'Oro}, A. and {Esquej}, P. and {Fern{\'a}ndez-Hern{\'a}ndez}, J. and {Fraile}, E. and {Garabato}, D. and {Garc{\'\i}a-Lario}, P. and {Gosset}, E. and {Haigron}, R. and {Halbwachs}, J.-L. and {Hambly}, N.~C. and {Harrison}, D.~L. and {Hern{\'a}ndez}, J. and {Hestroffer}, D. and {Hodgkin}, S.~T. and {Holl}, B. and {Jan{\ss}en}, K. and {Jevardat de Fombelle}, G. and {Jordan}, S. and {Krone-Martins}, A. and {Lanzafame}, A.~C. and {L{\"o}ffler}, W. and {Marchal}, O. and {Marrese}, P.~M. and {Moitinho}, A. and {Muinonen}, K. and {Osborne}, P. and {Pancino}, E. and {Pauwels}, T. and {Recio-Blanco}, A. and {Reyl{\'e}}, C. and {Riello}, M. and {Rimoldini}, L. and {Roegiers}, T. and {Rybizki}, J. and {Sarro}, L.~M. and {Siopis}, C. and {Smith}, M. and {Sozzetti}, A. and {Utrilla}, E. and {van Leeuwen}, M. and {Abbas}, U. and {{\'A}brah{\'a}m}, P. and {Abreu Aramburu}, A. and {Aerts}, C. and {Aguado}, J.~J. and {Ajaj}, M. and {Aldea-Montero}, F. and {Altavilla}, G. and {{\'A}lvarez}, M.~A. and {Alves}, J. and {Anders}, F. and {Anderson}, R.~I. and {Anglada Varela}, E. and {Antoja}, T. and {Baines}, D. and {Baker}, S.~G. and {Balaguer-N{\'u}{\~n}ez}, L. and {Balbinot}, E. and {Balog}, Z. and {Barache}, C. and {Barbato}, D. and {Barros}, M. and {Barstow}, M.~A. and {Bartolom{\'e}}, S. and {Bassilana}, J.-L. and {Bauchet}, N. and {Becciani}, U. and {Bellazzini}, M. and {Berihuete}, A. and {Bernet}, M. and {Bertone}, S. and {Bianchi}, L. and {Binnenfeld}, A. and {Blanco-Cuaresma}, S. and {Blazere}, A. and {Boch}, T. and {Bombrun}, A. and {Bossini}, D. and {Bouquillon}, S. and {Bragaglia}, A. and {Bramante}, L. and {Breedt}, E. and {Bressan}, A. and {Brouillet}, N. and {Brugaletta}, E. and {Bucciarelli}, B. and {Burlacu}, A. and {Butkevich}, A.~G. and {Buzzi}, R. and {Caffau}, E. and {Cancelliere}, R. and {Cantat-Gaudin}, T. and {Carballo}, R. and {Carlucci}, T. and {Carnerero}, M.~I. and {Carrasco}, J.~M. and {Casamiquela}, L. and {Castellani}, M. and {Castro-Ginard}, A. and {Chaoul}, L. and {Charlot}, P. and {Chemin}, L. and {Chiaramida}, V. and {Chiavassa}, A. and {Chornay}, N. and {Comoretto}, G. and {Contursi}, G. and {Cooper}, W.~J. and {Cornez}, T. and {Cowell}, S. and {Crifo}, F. and {Cropper}, M. and {Crosta}, M. and {Crowley}, C. and {Dafonte}, C. and {Dapergolas}, A. and {David}, M. and {David}, P. and {de Laverny}, P. and {De Luise}, F. and {De March}, R.},
        title = "{Gaia Data Release 3. Summary of the content and survey properties}",
      journal = {\aap},
         year = 2023,
        month = jun,
       volume = {674},
          eid = {A1},
        pages = {A1},
          doi = {10.1051/0004-6361/202243940},
archivePrefix = {arXiv},
       eprint = {2208.00211},
 primaryClass = {astro-ph.GA},
       adsurl = {https://ui.adsabs.harvard.edu/abs/2023A&A...674A...1G}
}

@ARTICLE{2026A&A...713A.177M,
       author = {{Ma{\'\i}z Apell{\'a}niz}, J. and {Gamen}, R.~C. and {Holgado}, G. and {Rosu}, S. and {Arias}, J.~I. and {Sim{\'o}n-D{\'\i}az}, S. and {Pellerin}, A. and {Abdul-Masih}, M. and {Madero Fuentes}, E. and {Molina-Calzada}, J.~A. and {Barb{\'a}}, R.~H.},
        title = "{Multiplicity of Massive stars in the Milky Way (M$^{3}$W): I. Project description, UNWIND, application to GLS 11 448 (=LS III +46 11), and DIB catalog}",
      journal = {\aap},
         year = 2026,
        month = sep,
       volume = {713},
          eid = {A177},
        pages = {A177},
          doi = {10.1051/0004-6361/202660211},
archivePrefix = {arXiv},
       eprint = {2604.02111},
 primaryClass = {astro-ph.SR},
       adsurl = {https://ui.adsabs.harvard.edu/abs/2026A&A...713A.177M}
}

@ARTICLE{2026AJ....171..256L,
       author = {{Liu}, Qian and {Bai}, Zhongrui and {Zhou}, Ming and {Yang}, Mingkuan and {Yang}, Xiaozhen and {Jiang}, Ziyue and {Yuan}, Hailong and {Li}, Ganyu and {He}, Yuji and {Wang}, Mengxin and {Dong}, Yiqiao and {Zhang}, Haotong},
        title = "{Characterizing the Instrumental Profile of LAMOST}",
      journal = {\aj},
         year = 2026,
        month = apr,
       volume = {171},
       number = {4},
          eid = {256},
        pages = {256},
          doi = {10.3847/1538-3881/ae4ebe},
archivePrefix = {arXiv},
       eprint = {2603.09178},
 primaryClass = {astro-ph.IM},
       adsurl = {https://ui.adsabs.harvard.edu/abs/2026AJ....171..256L}
}

@ARTICLE{1996MNRAS.283L.105K,
       author = {{Kerr}, T.~H. and {Hibbins}, R.~E. and {Miles}, J.~R. and {Fossey}, S.~J. and {Somerville}, W.~B. and {Sarre}, P.~J.},
        title = "{Molecular rotational contour fitting of ultra-high-resolution profiles of diffuse interstellar bands}",
      journal = {\mnras},
         year = 1996,
        month = dec,
       volume = {283},
       number = {4},
        pages = {L105-L109},
          doi = {10.1093/mnras/283.4.L105},
       adsurl = {https://ui.adsabs.harvard.edu/abs/1996MNRAS.283L.105K}
}

@ARTICLE{2002A&A...384..215G,
       author = {{Galazutdinov}, G. and {Moutou}, C. and {Musaev}, F. and {Kre{\l}owski}, J.},
        title = "{Very high resolution profiles of 6196 {\r{A}}and 6614 {\r{A}}diffuse interstellar bands}",
      journal = {\aap},
         year = 2002,
        month = mar,
       volume = {384},
        pages = {215-224},
          doi = {10.1051/0004-6361:20020003},
       adsurl = {https://ui.adsabs.harvard.edu/abs/2002A&A...384..215G}
}

@ARTICLE{2002A&A...396..987G,
       author = {{Galazutdinov}, G. and {Stachowska}, W. and {Musaev}, F. and {Moutou}, C. and {Lo Curto}, G. and {Krelowski}, J.},
        title = "{Fine structure of profiles of weak diffuse interstellar bands}",
      journal = {\aap},
         year = 2002,
        month = dec,
       volume = {396},
        pages = {987-991},
          doi = {10.1051/0004-6361:20021299},
       adsurl = {https://ui.adsabs.harvard.edu/abs/2002A&A...396..987G}
}

@ARTICLE{2016A&A...590A..52O,
       author = {{Omont}, Alain},
        title = "{Interstellar fullerene compounds and diffuse interstellar bands}",
      journal = {\aap},
         year = 2016,
        month = may,
       volume = {590},
          eid = {A52},
        pages = {A52},
          doi = {10.1051/0004-6361/201527685},
archivePrefix = {arXiv},
       eprint = {1511.01250},
 primaryClass = {astro-ph.GA},
       adsurl = {https://ui.adsabs.harvard.edu/abs/2016A&A...590A..52O}
}

@ARTICLE{2002ApJ...578..877S,
       author = {{Snow}, Theodore P. and {Zukowski}, Daniel and {Massey}, Philip},
        title = "{The Intrinsic Profile of the 4428 {\r{A}} Diffuse Interstellar Band}",
      journal = {\apj},
         year = 2002,
        month = oct,
       volume = {578},
       number = {2},
        pages = {877-884},
          doi = {10.1086/342660},
       adsurl = {https://ui.adsabs.harvard.edu/abs/2002ApJ...578..877S}
}

@ARTICLE{2015Natur.523..322C,
       author = {{Campbell}, E.~K. and {Holz}, M. and {Gerlich}, D. and {Maier}, J.~P.},
        title = "{Laboratory confirmation of C$_{60}$$^{+}$ as the carrier of two diffuse interstellar bands}",
      journal = {\nat},
         year = 2015,
        month = jul,
       volume = {523},
       number = {7560},
        pages = {322-323},
          doi = {10.1038/nature14566},
       adsurl = {https://ui.adsabs.harvard.edu/abs/2015Natur.523..322C}
}

@ARTICLE{2022RAA....22h5007W,
       author = {{Wu}, Ke-Fei and {Luo}, A.-Li and {Chen}, Jian-Jun and {Hou}, Wen and {Zhao}, Yong-Heng},
        title = "{Diffuse Interstellar Bands {\ensuremath{\lambda}}6379, {\ensuremath{\lambda}}6614, and {\ensuremath{\lambda}}6660 in the LAMOST-MRS Spectra}",
      journal = {Research in Astronomy and Astrophysics},
         year = 2022,
        month = aug,
       volume = {22},
       number = {8},
          eid = {085007},
        pages = {085007},
          doi = {10.1088/1674-4527/ac7388},
       adsurl = {https://ui.adsabs.harvard.edu/abs/2022RAA....22h5007W}
}

@ARTICLE{2015MNRAS.452.3629L,
       author = {{Lan}, Ting-Wen and {M{\'e}nard}, Brice and {Zhu}, Guangtun},
        title = "{Exploring the diffuse interstellar bands with the Sloan Digital Sky Survey}",
      journal = {\mnras},
         year = 2015,
        month = oct,
       volume = {452},
       number = {4},
        pages = {3629-3649},
          doi = {10.1093/mnras/stv1519},
archivePrefix = {arXiv},
       eprint = {1406.7284},
 primaryClass = {astro-ph.GA},
       adsurl = {https://ui.adsabs.harvard.edu/abs/2015MNRAS.452.3629L}
}

@ARTICLE{2024A&A...691A.282M,
       author = {{Ma}, Xiao-Xiao and {Chen}, Jian-Jun and {Luo}, A.-Li and {Zhao}, He and {Shi}, Ji-Wei and {Chen}, Jing and {Liang}, Jun-Chao and {Ma}, Shu-Guo and {Qu}, Cai-Xia and {Jiang}, Bi-Wei},
        title = "{Measuring the diffuse interstellar bands at 5780, 5797, and 6614 {\r{A}} in low-resolution spectra of cool stars from LAMOST}",
      journal = {\aap},
         year = 2024,
        month = nov,
       volume = {691},
          eid = {A282},
        pages = {A282},
          doi = {10.1051/0004-6361/202451408},
archivePrefix = {arXiv},
       eprint = {2409.19539},
 primaryClass = {astro-ph.GA},
       adsurl = {https://ui.adsabs.harvard.edu/abs/2024A&A...691A.282M}
}

@ARTICLE{2013ApJ...778...86K,
       author = {{Kos}, J. and {Zwitter}, T. and {Grebel}, E.~K. and {Bienayme}, O. and {Binney}, J. and {Bland-Hawthorn}, J. and {Freeman}, K.~C. and {Gibson}, B.~K. and {Gilmore}, G. and {Kordopatis}, G. and {Navarro}, J.~F. and {Parker}, Q. and {Reid}, W.~A. and {Seabroke}, G. and {Siebert}, A. and {Siviero}, A. and {Steinmetz}, M. and {Watson}, F. and {Wyse}, R.~F.~G.},
        title = "{Diffuse Interstellar Band at 8620 {\r{A}} in RAVE: A New Method for Detecting the Diffuse Interstellar Band in Spectra of Cool Stars}",
      journal = {\apj},
         year = 2013,
        month = dec,
       volume = {778},
       number = {2},
          eid = {86},
        pages = {86},
          doi = {10.1088/0004-637X/778/2/86},
archivePrefix = {arXiv},
       eprint = {1309.4273},
 primaryClass = {astro-ph.GA},
       adsurl = {https://ui.adsabs.harvard.edu/abs/2013ApJ...778...86K}
}

@ARTICLE{2012MNRAS.425.1763Y,
       author = {{Yuan}, H.~B. and {Liu}, X.~W.},
        title = "{Detections of diffuse interstellar bands in the SDSS low-resolution spectra}",
      journal = {\mnras},
         year = 2012,
        month = sep,
       volume = {425},
       number = {3},
        pages = {1763-1771},
          doi = {10.1111/j.1365-2966.2012.21674.x},
archivePrefix = {arXiv},
       eprint = {1207.1573},
 primaryClass = {astro-ph.SR},
       adsurl = {https://ui.adsabs.harvard.edu/abs/2012MNRAS.425.1763Y}
}

\label{lastpage}

\end{document}